\PassOptionsToPackage{backref=page}{hyperref}
\documentclass[AMA,STIX1COL]{WileyNJD-v2}

\usepackage[framemethod=TikZ]{mdframed}
\usepackage{amsmath,amsfonts,listing,color,graphicx,wrapfig, graphicx}
\usepackage{subcaption, comment}
\usepackage{tikz}
\usetikzlibrary{positioning}
\usetikzlibrary[shapes.misc]
\usetikzlibrary{arrows,automata}
\usepackage{todonotes}
\DeclareUnicodeCharacter{2212}{-}
\articletype{Research article}
\received{XXXX-XX-XX}
\revised{XXXX-XX-XX}
\accepted{XXXX-XX-XX}

\begin{document}

\lstset{ %
  language=R,                           
  basicstyle=\footnotesize\ttfamily,    
  numbers=left,                         
  numberstyle=\tiny\color{gray},        
  stepnumber=1,                         
  numbersep=5pt,                        
  backgroundcolor=\color{white},        
  showspaces=false,                     
  showstringspaces=false,               
  showtabs=false,                       
  frame=single,                         
  rulecolor=\color{black},              
  tabsize=1,                            
  captionpos=b,                         
  breaklines=true,                      
  inputencoding={utf8},extendedchars=false, 
  breakatwhitespace=false,              
  title=\lstname,                       
  keywordstyle=\color{blue},            
  commentstyle=\color{green!60!black},  
  stringstyle=\color{purple!60!black},  
  escapeinside={\?***}{***\?)}
}

\title{The Delta-Method and Influence Function in Medical Statistics: a Reproducible Tutorial}

\author[b]{Rodrigo Zepeda-Tello}
\author[c,d]{Michael Schomaker}
\author[e]{Camille Maringe}
\author[e]{Matthew J. Smith}
\author[e]{Aurelien Belot}
\author[e]{Bernard Rachet}
\author[f,g]{Mireille E. Schnitzer}
\author[e,h]{Miguel Angel Luque Fernandez*}

\authormark{Zepeda-Tello, Luque-Fernandez \emph{et al}.}

\address[b]{Dirección de Prestaciones Económicas y Sociales, Instituto Mexicano del Seguro Social, Mexico City, Mexico}
\address[c]{Institut für Statistik, Ludwig-Maximilians-Universität München, 80799 München, Germany}
\address[d]{Centre for Infectious Disease Epidemiology and Research, University of Cape Town, South Africa}
\address[e]{ICON-group. Non-communicable Disease Epidemiology. London School of Hygiene and
Tropical Medicine. London, U.K.}
\address[f]{Faculty of Pharmacy and Department of Social and Preventive Medicine, Universit\'e de Montr\'eal, Montreal, Canada}
\address[g]{Department of Epidemiology, Biostatistics and Occupational Health, McGill University, Montreal, Canada}
\address[h]{Department of Statistics and Operations Research, University of Granada, Granada, Spain.}

\corres{Miguel Angel Luque Fernandez
\email{miguel-angel.luque@lshtm.ac.uk}}

\presentaddress{Keppel St, London WC1E 7HT, UK}

\abstract{\textbf{ABSTRACT}
Approximate statistical inference via determination of the asymptotic distribution of a statistic is routinely used for inference in applied medical statistics (e.g. to estimate the standard error of the marginal or conditional risk ratio). One method for variance estimation is the classical Delta-method but there is a knowledge gap as this method is not routinely included in training for applied medical statistics and its uses are not widely understood. Given that a smooth function of an asymptotically normal estimator is also asymptotically normally distributed, the Delta-method allows approximating the large-sample variance of a function of an estimator with known large-sample properties. In a more general setting, it is a technique for approximating the variance of a functional (i.e., an estimand) that takes a function as an input and applies another function to it (e.g. the expectation function). Specifically, we may approximate the variance of the function using the functional Delta-method based on the influence function (IF). The IF explores how a functional $\phi(\theta)$ changes in response to small perturbations in the sample distribution of the estimator and allows to compute the empirical standard error of the distribution of the functional. The ongoing development of new methods and techniques may pose a challenge for applied statisticians who are interested in mastering the application of these methods. In this tutorial we review the use of the classical and functional Delta-method and their links to the IF from a practical perspective. We illustrate the methods using a cancer epidemiology example and we provide reproducible and commented code in R and Python using symbolic programming. The code can be accessed at \hyperlink{https://github.com/migariane/DeltaMethodInfluenceFunction}{https://github.com/migariane/DeltaMethodInfluenceFunction}
}

\keywords{Statistical Inference; Influence Function; Delta-method; Epidemiology; Tutorial}

\maketitle
\section{Introduction}

\noindent A fundamental problem in inferential statistics is to approximate the distribution of an estimator constructed from the sample (\textit{i.e.} a statistic). The standard error (SE) of an estimator characterises its variability.\citep{Boos2013} Oftentimes, it is not directly the estimator which is of interest but a function of it. In this case, the Delta-Method can approximate the standard error (with known asymptotic properties) using Taylor expansions because a smooth function of an asymptotically normal estimator is also asymptotically normal. \citep{Vaart1998} In a more general setting, this technique is also useful for approximating the variance of some functionals. For instance, in epidemiology the Delta-method is used to compute the SE of functions such as the risk difference (RD) and the risk ratio (RR),\citep{Agresti2010} which are all functions of the risk (a parameter representing the probability of the outcome).\citep{Armitage2005, Boos2013} Alternatively to the  Delta-method to approximate the distribution of the SE \citep{Boos2013, MillarMaximumADMB} for large samples, we can use other computational methods such as the bootstrap.\citep{Efron1993, efron1982} In the course of their research, it may be necessary for applied statisticians to assess whether a large sample approximation of the distribution of a statistic is appropriate, how to derive the approximation, and how to use it for inference in applications. The distribution of the statistic must be approximated to directly estimate its variance and hence the SE because the number and type of inference problems for which it can be analytically determined is narrow.

In this tutorial we introduce the use of the classical and functional Delta-method, the Influence Function (IF), and their relationship from a practical perspective. Hampel introduced the concept of the IF in 1974.\citep{hampel1974} He highlighted that most estimators can actually be viewed as functionals constructed from the distribution functions. The IF was further developed in the context of robust statistics but is now used in many fields, including causal inference.\citep{hampel1974} The IF is often used to approximate the SE of a plug-in asymptotically linear estimator.\citep{Tsiatis:2007aa} Mathematically, the IF is derived using the second term of the first order Taylor expansion used to empirically approximate the distribution of the plug-in estimator.\citep{Boos2013} It can be easily derived for most common estimators and it appears in the formulas for asymptotic variances of asymptotically normally distributed estimators. The IF is equivalent to the normalized score functions of maximum likelihood estimators.\citep{hampel1974}

Furthermore, the tutorial includes boxes with R code (R Foundation for Statistical Computing, Vienna, Austria)\citep{R2020} to support the implementation of the methods and to allow readers to learn by doing. The code can be accessed at \hyperlink{https://github.com/migariane/DeltaMethodInfluenceFunction}{https://github.com/migariane/DeltaMethodInfluenceFunction}. In section 1, we introduce the importance of the Delta-method in statistics and justify the need of a tutorial for applied statisticians. In section 2, we review the theory of the classical and functional Delta-methods and the influence function (IF). In section 3, we provide multiple worked examples and code for applications of the classical and functional Delta-method, and the IF. The first examples involve deriving the SE for the sample mean of a variable, the ratio of two means of two independent variables, and the ratio of two sample proportions (i.e. the risk ratio). Also, we provide a  example where the required conditions for the for the Delta-method do not hold. We then show how to use the functional Delta-method based on the IF to derive the SE for the quantile function and the correlation coefficient. Our final example is motivated by an application in cancer epidemiology and involves a parameter of interest that is a combination of coefficients in a logistic regression model. Finally, in section 4, we provide a concise conclusion where we mention additional interconnected methods with the Delta-method and the IF such as M-estimation and the Huber Sandwich estimator.

\section{Theory: The Classical Delta-method}

Let $\theta$ be a parameter. For this tutorial, we are interested in working with an estimand $\psi$ that can be written as a function of $\theta$ (\textit{i.e.,}  $\psi := \phi(\theta)$) rather than $\theta$ itself. For instance, we may not be interested in the probability of having a particular disease, but in the ratio of two probabilities $\theta=(\theta_1,\theta_2)$, where the first probability ($\theta_1$) is of developing the disease under treatment, the second ($\theta_2$) is of developing the disease without treatment. The estimand $\psi  = \phi(\theta)=\theta_1/\theta_2$ represents the relative risk. Define the estimator of $\psi$ to be $\hat{\psi}=\hat{\theta}_1/\hat{\theta}_2$, the ratio of the estimators of the respective probabilities. The question is: if we know the variances of $\hat{\theta}_1$ and $\hat{\theta}_2$, \textit{how do we obtain the variance of $\hat{\psi}$?} The Delta-method is one approach to answer this.

Let $\hat{\psi} := \phi(\hat{\theta})$ be an estimator of  $\psi = \phi(\theta)$ from a random sample $X_1, X_2, \dots, X_n$ where the $X_i$s are independent and identically distributed (i.i.d) with a distribution defined with a parameter $\theta$ (\textit{i.e.} $X_i \sim F_{X_i}(\cdot | \theta)$). Examples of parameters include the rate of an exponential variable ($\theta = \lambda$), the mean and variance of a normal distribution $\theta = (\mu, \sigma^2)$ or the probability of a specific category under a multinomial model with $m$ different categories: $\theta = (p_1, p_2, \dots, p_m)^T$ with $\sum_{j=1}^m p_j = 1$.

Any (measurable) function $T_n := T(X_1, \dots, X_n)$ of the random sample is called a statistic.\citep{Casella1998TheoryEstimation} In particular, any estimator $\hat{\theta}$ of $\theta$ is a function of the random sample making it a statistic. For example, if $\theta = \mu$, the mean, $\hat{\theta}(X_1, \dots, X_n) = 1/n \sum_{i =1}^n X_i = \bar{X}$ is a function of the $X_i$s. To emphasize the dependency of the estimator, $\hat{\theta}$, on the sample size, $n$, we write: $\hat{\theta}_n$. Thus $\hat{\theta}_{100}$ would denote the estimator under a random sample of size $100$ and $\hat{\theta}_{\infty} := \lim_{n \to \infty} \hat{\theta}_n$ denotes the estimator under a  random sample ``of infinite size''. Any (measurable) function of the estimator, $\hat{\psi}_n := \phi(\hat{\theta}_n)$ also depends upon the random sample and hence it is a statistic too. Due to the dependency upon the random sample, any statistic by itself is a random variable. We can thus characterise the estimator in terms of its distribution. As an example, if the $X_i$ are i.i.d. $\textrm{Normal}(\mu,\sigma^2)$ then $\hat{\theta}_n =\bar{X}$ also has a normal distribution with parameters $(\mu,\sigma^2/n)$. Furthermore, the statistic $\hat{\psi}_n = \psi(\hat{\theta}_n) = e^{\hat{\theta}_n}$ has a $\textrm{LogNormal}(\mu, \sigma^2/n)$ distribution. 

More often than not, the distribution of a statistic cannot be estimated directly and we rely on the asymptotic (large sample) properties of $\hat{\theta}_{n}$ where $n$ approaches $\infty$. A most powerful and well-known result is the central limit theorem which states under reasonable regularity conditions (i.i.d. variables with mean $\mu$ and standard deviation $\sigma$)\citep{Billingsley1961StatisticalChains} that if $\hat{\theta}_n =  \bar{X}$ then, for large $n$,
\begin{equation}
 \sqrt{n}\big(\hat{\theta}_{n} - \mu\big) \overset{\text{approx}}{\sim} \textrm{Normal}(0, \sigma^2)   
\end{equation}
which is the property that allows us to construct the Wald-type asymptotic confidence intervals: $\hat{\theta}_n \pm Z_{1 - \alpha/2}\cdot \sqrt{\sigma^2/n}$.\citep{Agresti2012ApproximateProportions} 
\\

However, when the function $\phi(\cdot)$ -- a function of one or more estimators with large-sample normality with known variance -- is not linear (\textit{e.g.} the ratio of two proportions) and there is not a closed functional form to derive the SE, we use the Delta-method. The classical Delta-method states that under certain regularity conditions for the function $\phi(\cdot)$, the statistic $\hat{\theta}$, and the i.i.d. random variables $X_i$s, the distribution of $\phi(\hat{\theta})$ can be approximated via a normal distribution with a variance proportional to $\phi$'s rate of change at $\theta$, the derivative $\phi'(\theta)$. In the one dimensional case of $\theta\in\mathbb{R}$ and $\phi(\theta)\in\mathbb{R}$, if $\hat{\theta}_n$ is asymptotically normal, this theorem states that, for large $n$ (Appendix: Delta-method proof):
$$
\sqrt{n}\big(\phi(\hat{\theta}_{n}) - \phi(\mu)\big) \overset{\text{approx}}{\sim} \textrm{Normal}(0, \phi'(\theta)\sigma^2).
$$
This provides the researcher with  confidence intervals based on asymptotic normality:
$$
\hat{\theta}_n \pm Z_{1 - \alpha/2}\cdot \sqrt{\frac{\phi'(\theta)\sigma^2}{n}}.
$$
To better understand the Delta-method we need to discuss four concepts. First, we need to discuss how derivatives approximate functions such as $\phi$ via a Taylor expansion. Second, we describe convergence in distribution which is what allows us to characterise the asymptotic properties of the estimator. Third, we present the central limit theorem, which is at the core of the Delta-method. Finally we'll generalize these results to the functional Delta-method using influence functions.

\subsection{Taylor's Approximation}

\begin{figure}[!htb]
\begin{subfigure}{.475\textwidth}
  \centering
  \includegraphics[width=.9\linewidth]{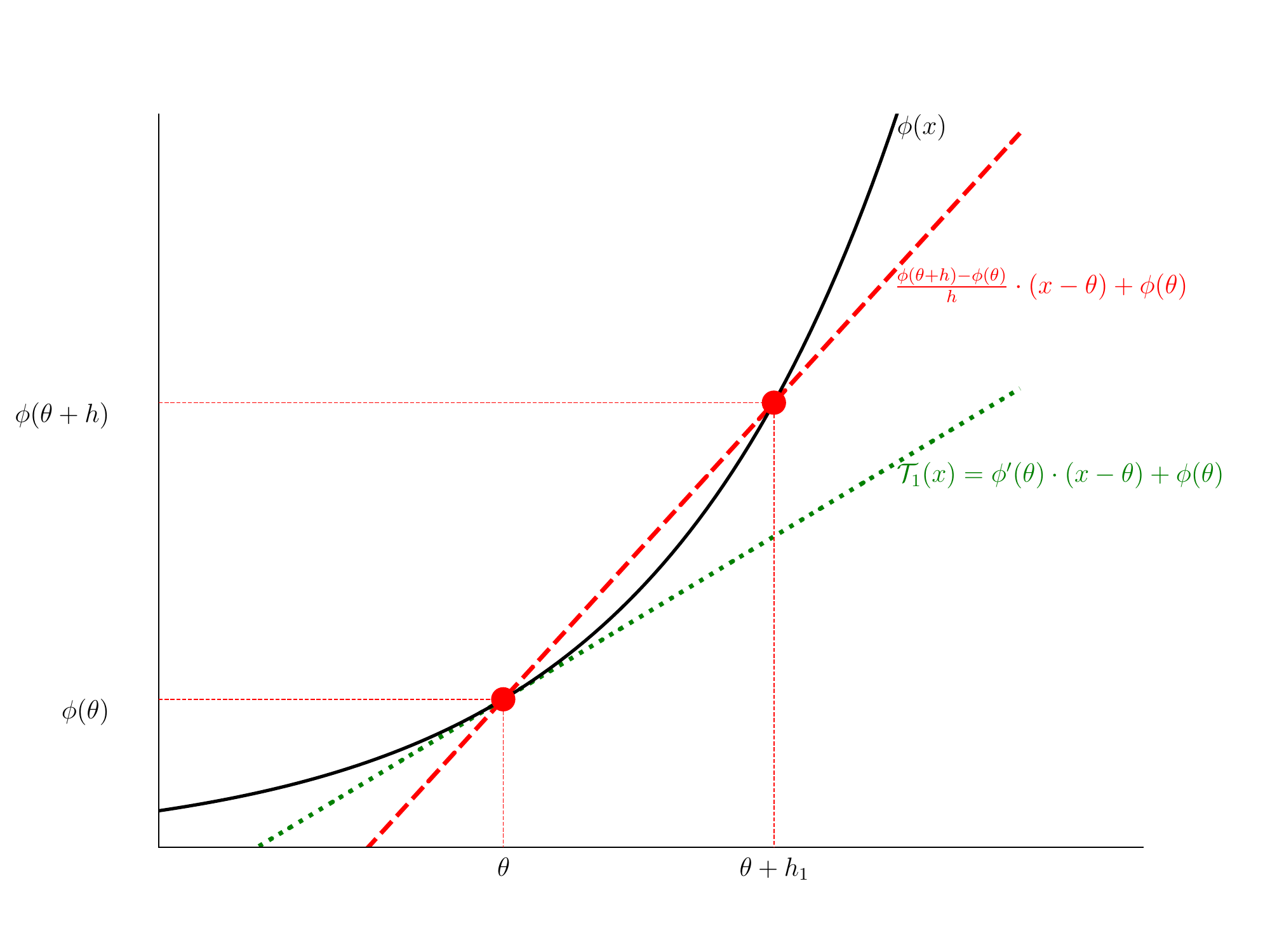}
  \caption{The derivative of the function $\phi(\theta) = e^{\theta}$ (black) corresponds to the slope of the tangent line $T_1$ (green) which is constructed by taking the line $\mathcal{L}(x) = \frac{\phi(\theta + h) - \phi(\theta)}{h}(x - \theta) + \phi(\theta)$ (red) and sending $h \to 0$.}
  \label{fig:sfig1}
\end{subfigure}\hfill
\begin{subfigure}{.475\textwidth}
  \centering
  \includegraphics[width=.9\linewidth]{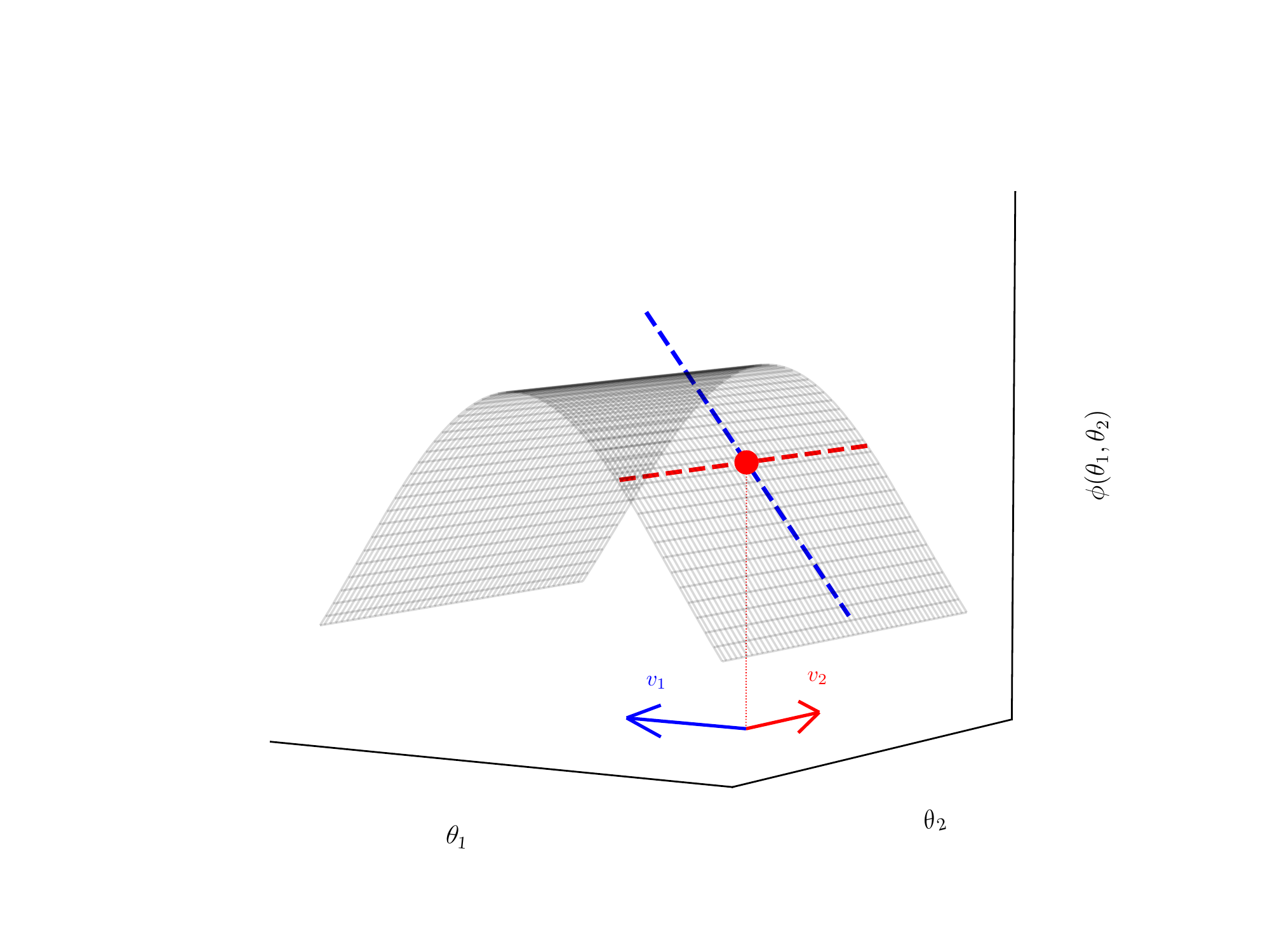}
  \caption{In a multidimensional setting infinite tangent lines are possible. Here we show the tangent lines to $\phi(\theta_1, \theta_2) = cos(\theta_1)^2$ in the directions of $v_1 = (1,0)^T$ (blue) and $v_2 = (0,1)^T$ (red).}
  \label{fig:sfig2}
\end{subfigure}
\vskip\baselineskip
\begin{subfigure}{.475\textwidth}
  \centering
  \includegraphics[width=.9\linewidth]{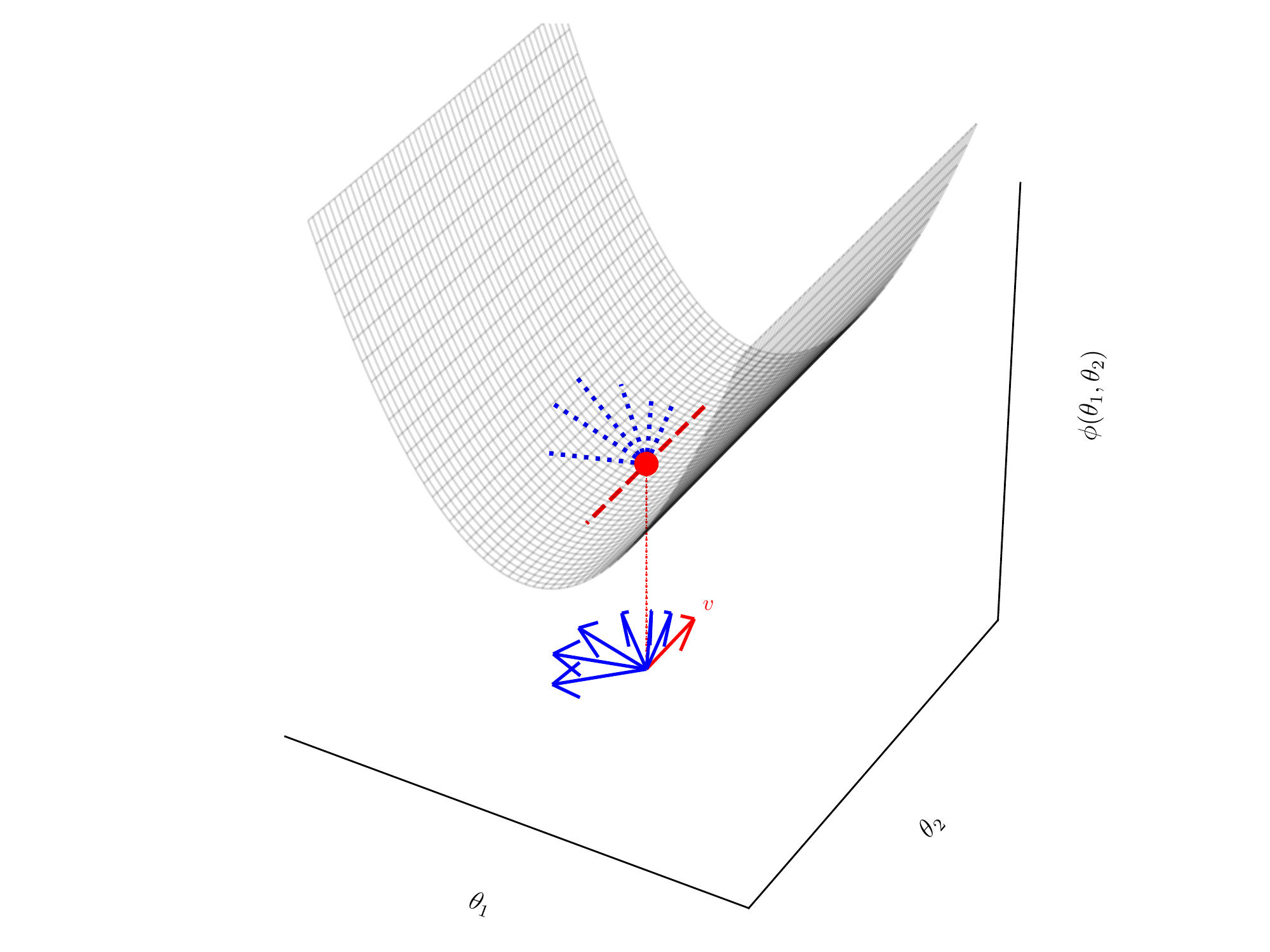}
  \caption{The Hadamard derivative is constructed via a sequence of directions $\{v_h\}$ (blue) such that $v_h \to v$ as $h \downarrow 0$ ($v$ is represented in red). The Hadamard derivative exists if this process is valid for any sequence $\{v_h\}$ that tends to $v$.}
  \label{fig:sfig3}
\end{subfigure}%
\hfill
\begin{subfigure}{.475\textwidth}
  \centering
  \includegraphics[width=.9\linewidth]{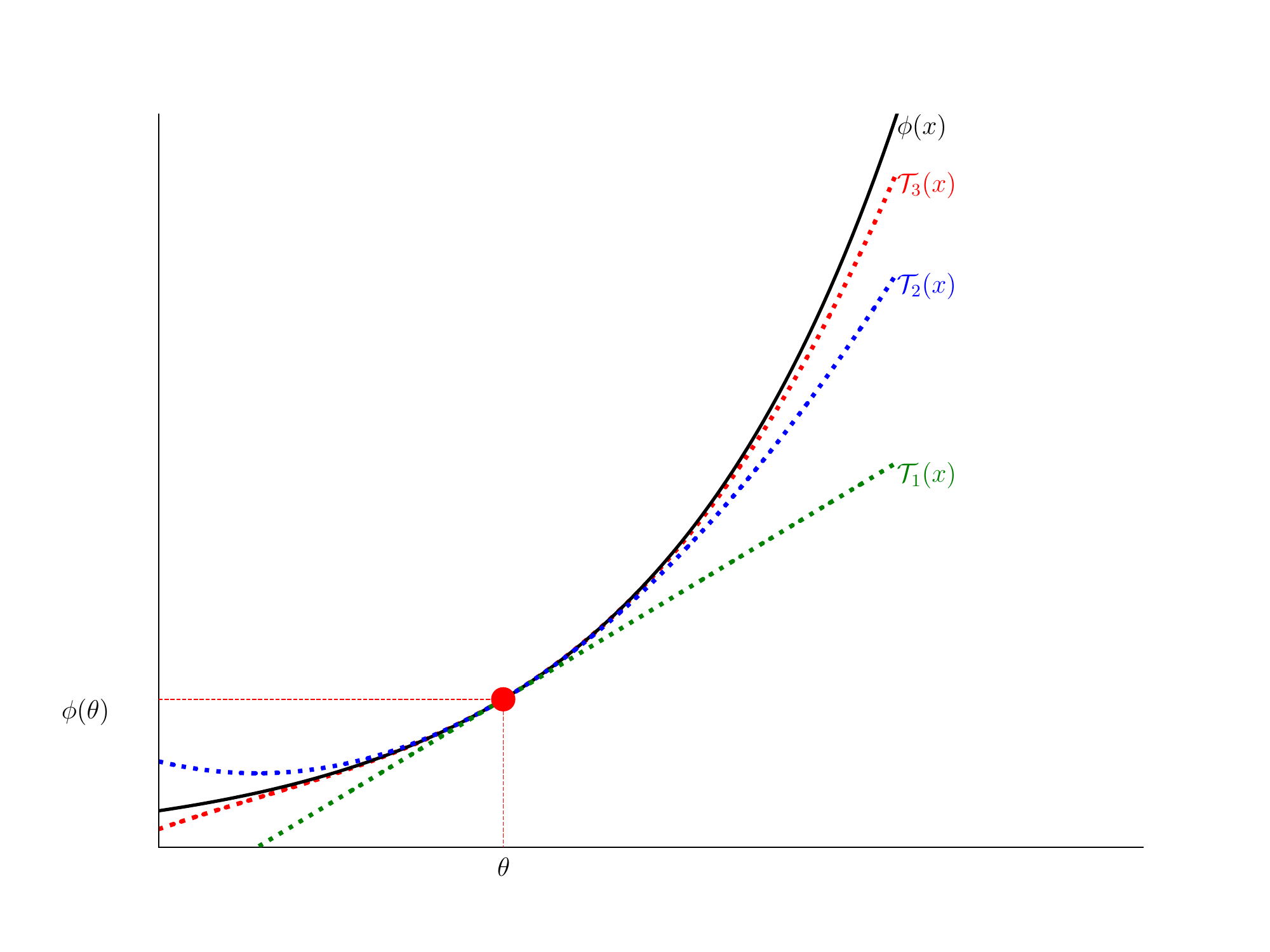}
  \caption{Taylor first $\mathcal{T}_1$, second $\mathcal{T}_2$ and third $\mathcal{T}_3$ order approximations to a function $\phi$. Intuitively, a higher order approximation results in a function closer to the true value.}
  \label{fig:sfig4}
\end{subfigure}
\caption{\normalsize. First derivative of the function $\phi(\theta) = e^{\theta}$ (a), tangent plane to $\phi(\theta_1, \theta_2) = cos(\theta_1)^2$ (b), Hadamard derivative (c) and Taylor approximation (d) representations.}
\label{fig:fig}
\end{figure}

For Taylor's approximation to work we need to have a function $\phi$ that is differentiable at $\theta$. Following the classical definition of differentiability,\citep{Courant1988DifferentialCalculus} a real valued function $\phi$ with domain $\mathbb{D}$, a subset of $\mathbb{R}$, ($\phi:\mathbb{D}\to\mathbb{R}$) is differentiable at $\theta\in\mathbb{D}$ and has derivative $\phi'(\theta)$ if the following limit exists:
$$
\phi'(\theta) := \lim_{\substack{h \to 0\\h \neq 0}} \dfrac{\phi(\theta + h) - \phi(\theta)}{h}.
$$
Intuitively, this definition states that one can estimate a unique tangent line to $\phi$ with slope $\phi'$ at $\theta$ by calculating the values of the function $\phi$ at $\theta$ and $\theta + h$ and reducing the size of $h$ (see Figure \ref{fig:sfig1}).

This definition can be extended to the multivariate case via directional derivatives (Gâteaux derivatives).\citep{Gateaux1919FonctionsIndependantes} In multiple dimensions, there is no one unique tangent line that can be generated (see Figure \ref{fig:sfig2}); hence, in addition to the function $\phi$, one must also specify the direction of the vector $v$ in which the tangent line will be calculated. This results in $\Tilde{\partial}_v \phi(\theta)$, the derivative of $\phi$ at $\theta$ in the direction $v$:\footnote{
You might notice a slight change in notation where the limit is stated as $h \downarrow 0$ instead of the classical $h \to 0$. The $\downarrow$ notation implies that the limit is taken with $h$ decreasing towards zero in order to distinguish the direction $v$ from $-v$.}

\begin{equation}
\Tilde{\partial}_v \phi(\theta) := \lim_{\substack{h \downarrow 0\\h \neq 0}} \dfrac{\phi(\theta + h\cdot v) - \phi(\theta)}{h}.
\end{equation}
As an example, Figure \ref{fig:sfig2} is a graph of the function $\phi(\theta_1, \theta_2) = \sin(\theta_1)$ with two different vectors $v_1 = (1,0)^T$ and $v_2 = (0, 1)^T$. Each vector results in a different directional derivative, $\Tilde{\partial}_{v_1} \phi (\theta) = \cos(\theta_1)$ and $\Tilde{\partial}_{v_2}\phi (\theta)= 0$ , respectively, corresponding to the slopes of the tangent lines in the directions of $v_1$ and $v_2$ respectively.

It turns out that for the Delta-method to be generalized to functionals (\textit{i.e.} functions of functions) having a Gâteaux derivative is not enough. We require not only that the directional derivative $\Tilde{\partial}_{v} \phi$ exists but also that it exists and coincides with the one obtained for any sequence of directions $\{ v_1, v_2, v_3, \dots\}$ that converge to $v$ (\textit{i.e.} $\lim_{h \downarrow 0} v_h = v$). This is called (equivalently) the compact derivative or the  Hadamard\citep{Beutner2016FunctionalFunctionals} (one-sided directional) \cite{Zajicek2014HadamardDifferentiability} derivative of $\phi$ at $\theta$ in the direction $v$ (as long as it is a linear function for any $v$) and is usually denoted as:
\begin{equation}\label{hadamard}
\partial_v \phi(\theta) := \lim_{\substack{h \downarrow 0\\h \neq 0}} \dfrac{\phi(\theta + h\cdot v_h) - \phi(\theta)}{h} \quad \text{ for any sequence } v_h \to v \text{ as } h \downarrow 0.
\end{equation}
This concept is illustrated in Figure \ref{fig:sfig3} where the specific sequence $v_h$ converges to $v$.

An equivalent definition of the Hadamard (one-sided directional) derivative which is useful for calculations involves setting $v = G - \theta$ for some function $G$ and $v_h = G_h - \theta$ with $G_h \to G$ which allows us to rewrite \eqref{hadamard} as:
\begin{equation}\label{hadamardos}
 \partial_{G - \theta} \phi(\theta) = \lim_{\substack{h \downarrow 0\\h \neq 0}} \dfrac{\phi\big((1 - h)\theta + h \cdot G_h \big) - \phi(\theta)}{h}\text{ for any sequence } G_h \to G \text{ as } h \downarrow 0.
\end{equation}

In the particular case of a constant sequence such that $G_h = G$ the expression reduces to a Gâteaux derivative which can oftentimes be computed as a classical derivative. We discuss a particular case of this derivative, the influence function, IF, (also known as influence curve) in Section \ref{if}. It is interpreted as the rate of change of our functional $\phi$ in the direction of a new observation, $x$.

Recall that the derivative, $\partial_v \phi(\theta)$, represents the slope of the line tangent to the function. Intuitively, if $\hat{\theta}$ is close to $\theta$, the tangent line at $\hat{\theta}$ should provide an adequate approximation of $\phi(\theta)$ Figure \ref{fig:sfig4}). This is stated in the  Taylor first order approximation of $\phi(\hat{\theta})$ around $\phi({\theta})$ as follows:
\begin{equation}\label{taylorhadamard}
\phi(\hat{\theta}_n) \approx \phi(\theta) + \partial_v \phi(\theta)
\end{equation}
with $v = \hat{\theta} - \theta$ and the sign $\approx$ is interpreted as \textit{approximately equal}. This can be rewritten as the more classical approach:

\begin{equation}\label{taylorhadamardnotation}
\phi(\hat{\theta}) - \phi(\theta) \approx  \partial_v \phi(\theta) \quad \text{ with } v = \hat{\theta}- \theta. 
\end{equation}

Readers might be familiar with the theorem in the classical notation of univariate calculus which states the approximation:
\begin{equation}\label{approxTaylorv1}
 \phi(\hat{\theta}) \approx \phi(\theta) + \phi'(\theta) \underbrace{(\hat{\theta} -  \theta)}_{v}.
\end{equation}
In this case the Hadamard derivative coincides with the classical one multiplied by $v = \hat{\theta} -  \theta$:
$$
\partial_v \phi(\theta) = \phi'(\theta) (\hat{\theta} -  \theta).
$$
The justification for this connection is given by Fréchet's derivative which represents the slope of the tangent plane. Intuitively, if the Hadamard (one-sided directional) derivatives $\partial_{v}\phi(\theta)$ exist for all directions $v$ we can talk about the tangent plane to $\phi$ at $\theta$. The tangent plane is ``made up'' of all the individual (infinite) tangent lines. The slope of the tangent plane is the Fréchet derivative $\nabla \phi$.\citep{Zajicek2014HadamardDifferentiability, ciarlet2013linear}. For univariate functions in $\phi:\mathbb{R}\to\mathbb{R}$ the Fréchet derivative is $\phi'$; for  functions of a multivariate $\theta$ returning one value, $\phi:\mathbb{R}^n\to\mathbb{R}$, this derivative is called the gradient and corresponds to the derivative of the function by each entry:
$$
\nabla \phi  = \Big( \frac{\partial \phi}{\partial \theta_1},\frac{\partial \phi}{\partial \theta_2}, \dots, \frac{\partial \phi}{\partial \theta_n} \Big).
$$
For multivariate functions, $\phi:\mathbb{R}^n\to\mathbb{R}^m$, the Fréchet derivative is an $m \times n$ matrix called the Jacobian (matrix):
\begin{equation}\label{jacobian}
\nabla \phi  = \begin{pmatrix}
 \frac{\partial \phi_1}{\partial \theta_1} & \frac{\partial \phi_1}{\partial \theta_2} & \dots & \frac{\partial \phi_1}{\partial \theta_n} \\
 \frac{\partial \phi_2}{\partial \theta_1} & \frac{\partial \phi_2}{\partial \theta_2} & \dots & \frac{\partial \phi_2}{\partial \theta_n} \\
 \vdots & \vdots & \ddots & \vdots \\
 \frac{\partial \phi_m}{\partial \theta_1} & \frac{\partial \phi_m}{\partial \theta_2} & \dots & \frac{\partial \phi_m}{\partial \theta_n} \\
\end{pmatrix}.
\end{equation}

To obtain the Hadamard (one-sided directional) derivative from the Fréchet derivatives, either $\phi'$ or $\nabla \phi$, one needs to apply the derivative operator $\nabla \phi$ to the direction vector $v$. This operation can be seen as ``projecting'' the tangent plane into the direction of $v$ hence resulting in the directional derivative: 
\begin{equation}\label{eqn:hadamard_from_frechet}
\partial_v \phi(\theta) = \phi'(\theta) \cdot v \quad \textrm{ or } \quad \partial_v \phi(\theta) = \nabla \phi(\theta)^T v.
\end{equation}

Thus the notation in \eqref{taylorhadamardnotation} which we'll use for the remainder of the paper includes not only the functional scenario but also the classical cases of functions in $\mathbb{R}$ and $\mathbb{R}^n$ respectively which can be obtained as the usual (classical) Fréchet derivatives projected onto $v$.

Finally, as a side note, we remark that it is possible to improve the approximation via higher order Taylor’s expansion around $\theta$  (see \ref{fig:sfig4})\citep{Courant1988DifferentialCalculus, ren2001second}:
$$
T_n(\hat{\theta}) = \phi(\theta) + \sum_{i = 1}^n \phi^{(n)}(\theta) \cdot \dfrac{(\theta - \hat{\theta})^n}{n!}
$$
where $\phi^{(n)}$ denotes the $n$-th derivative of $\phi$ defined as the derivative of the $(n-1)$-th derivative. Readers interested in pursuing higher order Hadamard derivatives can consult Ren and Sen (2001) and Tung and Bao (2022) \citep{REN2001187,tung2022higher}.

\subsection{Convergence in distribution}
For any random variable, $X$, the cumulative distribution function (CDF), also commonly referred to as the distribution function, quantifies the probability that $X$ is less than or equal to a real number $z$. Thus $X$'s (i.e., the CDF) is given by:
$$
F_X(z) = \mathbb{P}(X \leq z)
$$
where the sign $\leq$ is interpreted pointwise if $X$ is a random vector of size $n$ (\textit{i.e.} $X \leq z$ implies $X_1 \leq z_1$, $X_2 \leq z_2$, etc. for the vector $z = (z_1, \dots, z_n)$). The distribution function completely determines all the probabilities associated with a random variable as, for example, $\mathbb{P}(a < X \leq b)$ can be estimated as $F_X(b) - F_X(a)$ for any $a,b\in\mathbb{R}$.

Given a statistic $\hat{\theta}_n$ that depends upon the sample size, $n$, the statistic's distribution function also depends on $n$. Let $F_n$ denote the distribution of $\hat{\theta}_n$ and $F_{\Theta}$ be the distribution of a random variable, $\Theta$. We say that $\hat{\theta}_n$ \textbf{converges in distribution} to the random variable $\Theta$  if the CDF of $\Theta$ and the distribution of $\hat{\theta}_n$ coincide at infinity:
$$
\lim_{n \to \infty} F_n = F_{\Theta}.
$$
We remark that convergence in distribution does not imply that the random variables $\Theta$ and $\hat{\theta}_{\infty}$ are the same; it solely entails that the probabilistic model of $\Theta$ and $\hat{\theta}_{\infty}$  are identical (\textit{e.g.} both are $\text{Normal}(0,1)$) They are different random variables with a common distribution. Convergence in distribution is usually interpreted as an approximation stating that for large $n$, the distribution of $\hat{\theta}_n$ is approximately $F_{\Theta}$ (written $\hat{\theta}_n\overset{\text{approx}}{\sim}F_{\Theta}$).

\begin{figure}[!htb]
    \centering
    \includegraphics[width = 0.8\textwidth]{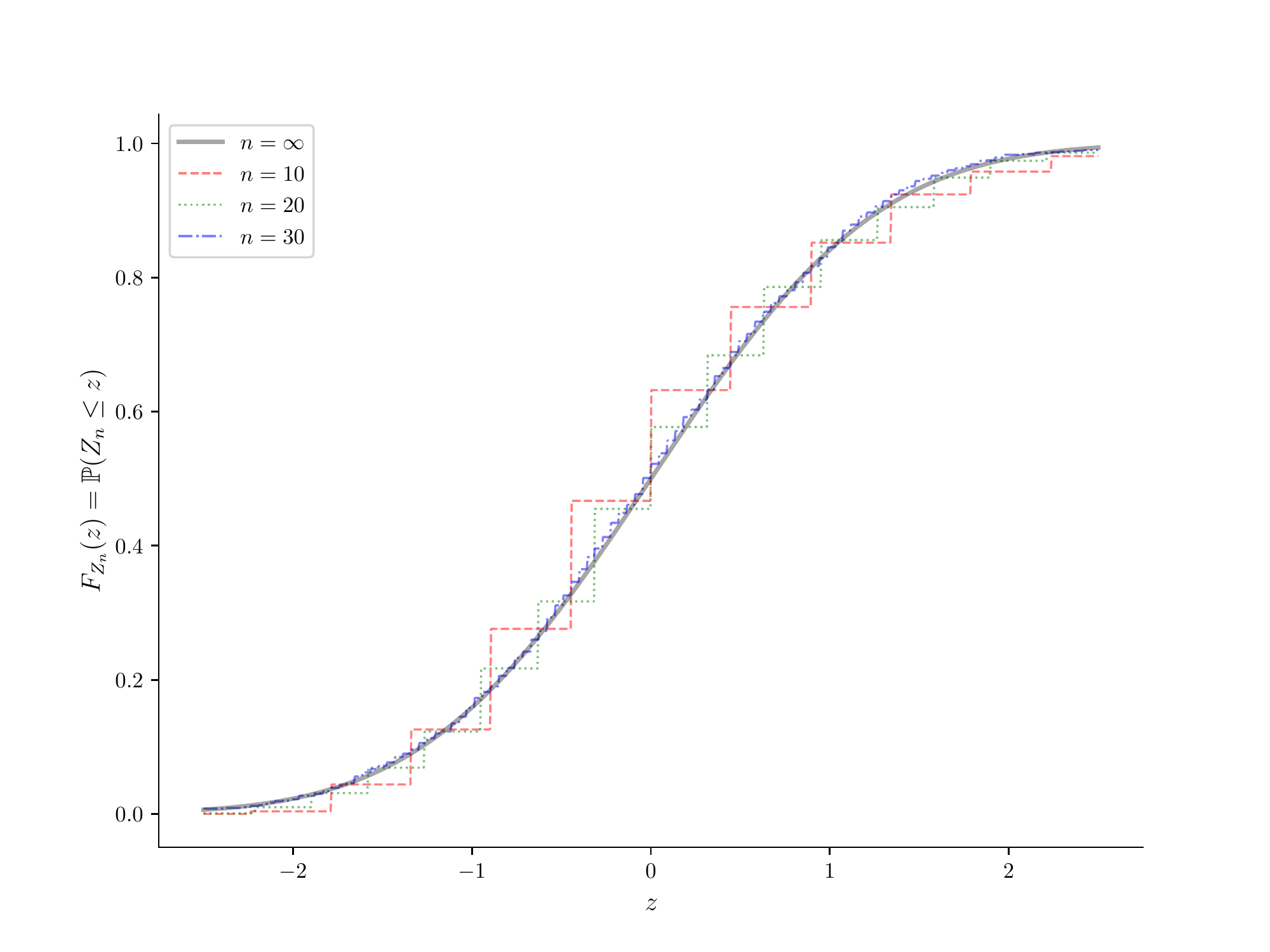}
    \caption{\normalsize. Convergence in distribution for the transformation $Z_n = \sqrt{n} \cdot \frac{\hat{\theta}_n - \lambda}{\sqrt{\lambda}}$ under different sample sizes, $n$, when the sample of $X_i$s comes from a Poisson distribution with parameter $\lambda = 1$ and $\hat{\theta}_n = \frac{1}{n} \sum_{i = 1}^n X_i$.}
    \label{fig:ecdf}
\end{figure}

One of the most important results concerning convergence in distribution is the Central Limit Theorem (CLT). The CLT applies to any random sample $\{X_1, X_2, \dots, X_n\}$ with $\mathbb{E}[X_i] = \mu$ and finite variance: $\text{Var}[X_i] = \sigma^2$. It states that the error of the sample mean, $\hat{\theta}_n = \frac{1}{n} \sum_{i = 1}^n X_i$,  times the square root of the sample size is normally distributed:
\begin{equation}\label{zconvergence}
\sqrt{\frac{n}{\sigma^2}}\Big( \hat{\theta}_n - \mu \Big) \overset{\text{d}}{\to} Z\,,
\end{equation}
where $Z \sim \textrm{Normal}(0,1)$ and $\overset{\text{d}}{\to}$ stands for convergence in distribution as $n \to \infty$. Figure \ref{fig:ecdf} illustrates the distribution of $Z_n = \sqrt{\frac{n}{\sigma^2}}\Big( \hat{\theta}_n - \mu \Big)$ for different sample sizes, $n$, when the  $X_i$s are $\text{Poisson}(1)$ distributed.

\subsection{Two sides of the same coin: the classical and functional Delta-method}
The Delta-method uses both the Taylor approximation and the concept of convergence in distribution. It states that if for some series of numbers that depend on the sample size, $\{r_n\}_{n = 1}^{\infty}$ with $\lim_{n \to \infty} r_n = \infty$, we have that $r_n(\hat{\theta}_{n} - \theta)$  converges in distribution to $Z$ then the weighted difference, $r_n\big(\phi(\hat{\theta}_{n}) - \phi(\theta)\big)$, converges to the distribution of the derivative of $\phi$ in the direction of $Z$:
$$
r_n\big(\phi(\hat{\theta}_{n}) - \phi(\theta)\big)\overset{\text{d}}{\to} \nabla \phi(\theta)^T Z,
$$
as long as $\phi$ is a function that can be approximated via its Taylor Series around $\theta$. Examples of numbers $r_n$ include $r_n = \sqrt{\frac{n}{\sigma^2}}$ as in \eqref{zconvergence}. The idea behind the Delta-method relates to the fact that we can transform \eqref{approxTaylorv1} into:
\begin{equation}\label{approxTaylorv2}
 r_n \Big( \phi(\hat{\theta}) - \phi(\theta) \Big) \approx  \nabla\phi(\theta)^T  r_n (\hat{\theta}_n -  \theta) \overset{\text{d}}{\to} \nabla \phi(\theta)^T Z
\end{equation}
where the random quantity, $r_n (\hat{\theta}_n -  \theta)$ converges in distribution to $Z$ and thus $r_n \big(\phi(\hat{\theta}_n) - \phi(\theta)\big)$ converges (approximately) to $\nabla\phi(\theta)^T Z$ (the derivative in the direction of $Z$). 

In practical terms this implies that the variance of $\phi(\hat{\theta}_n)$ can be approximated by an scaling of the variance of $\nabla\phi(\theta)^T Z$, \textit{i.e.}:
\begin{equation}\label{approxTaylorv3}
 \textrm{Var}\big[\phi(\hat{\theta}_n) - \phi(\theta)\big] = \textrm{Var}\big[\phi(\hat{\theta}_n)\big] \approx \frac{1}{r_n^2} \textrm{Var}\big[ \nabla\phi(\theta)^T Z\big].
\end{equation}

The same idea can be extended when the parameter of interest, $\theta$, is not a real number (or vector of numbers) but a function. In this case, $\phi$ is a functional (\textit{i.e.} a function of functions) and the corresponding method is oftentimes called \textbf{the functional Delta-method}. The result is that if $r_n(\hat{\theta}_{n} - \theta)\overset{\text{d}}{\to} Z$ with $Z$ now denoting a random function, then:
\begin{equation}\label{deltamethodfunctional}
r_n\big(\phi(\hat{\theta}_{n}) - \phi(\theta)\big)\overset{\text{d}}{\to} \partial_Z  \phi(\theta)
\end{equation}
where $\partial_Z \phi(\theta)$ denotes the Hadamard derivative of $\phi$ as  in \eqref{taylorhadamardnotation}. We remark that the theorem of \eqref{deltamethodfunctional} is general in the sense that it works for classical derivatives ($\partial_Z  \phi(\theta) = \phi'(\theta) Z$), gradients and jacobians ($\partial_Z  \phi(\theta) = \nabla \phi(\theta)^T Z$), and Hadamard derivatives ($\partial_Z  \phi(\theta) = \lim_{\substack{h \downarrow 0\\h \neq 0}} \big(\phi(\theta + h\cdot v_h) - \phi(\theta)\big) /h$) all following the notation from \eqref{hadamard}. 

The reader is invited to consult the supplementary material for the classical proof of the Delta-Method as well as the more general proof of the functional one.

\subsection{The influence function}\label{if}
It is common to represent scientific questions by \textit{estimands} (i.e., a quantity we are interested in estimating from our data). For example, suppose we are interested in a random variable $X$ which follows a (possibly unknown) discrete distribution $F_X(x)$. The variable $X$ might be a binary indicator for disease status, for example, in a particular population. If we are interested in the probability of having the given disease, our estimand is $\psi:= \mathbb{P}_X(X=1)$. In this case, we have $X \sim \textrm{Bernoulli}(\psi)$, the estimand is equivalent to the expectation of $X$, \textit{i.e.} $\psi := \mathbb{E}[X]$. The estimand can thus be seen as the parameter of the Bernoulli distribution. However a second interpretation is of importance: the estimand can also be seen as a \textbf{functional} as it takes a function -- specifically, the probability mass function $\mathbb{P}_X$ -- as an input and applies a function to it: the expectation. For $X$ taking discrete values, we have

\begin{equation}\label{pmf}
\psi = \phi \left( \mathbb{P}_X \right) = \sum_{x \in \mathcal{X}} x \cdot \mathbb{P}_X(X = x) = \mathbb{E}[X]
\end{equation}
where $\mathcal{X}$ denotes the support (\textit{i.e.} possible values) of $X$. In the binary case,  $\mathcal{X}=\{0,1\}$. If $X$ is continuous, an estimand defined as the expectation of $X$ is a functional of the probability density function $f(x)$,  such that $\mathbb{E}(X)=\int_{-\infty}^{\infty} x \cdot f(x)dx$.

It is important to highlight that the estimand $\psi$, which represents our scientific question, relates to a functional $\phi$ of the mass $\mathbb{P}_X$. Following the previous notation, we have that $\theta = \mathbb{P}_X$. If we have a random sample, $\{X_1,\dots,X_n\}$, we can compute the \textbf{empirical probability mass function (ePMF)}:
\begin{equation}\label{epmf}
\hat{\theta } = \hat{\mathbb{P}}_X(z) := \frac{1}{n} \sum\limits_{i = 1}^n \mathbb{I}_{\{X_i\}}(z) \qquad  \Bigg( = \frac{\text{Number of } X_i\text{s}  = z}{n} \Bigg)
\end{equation}
where the indicator function of a set $A$ is defined as
\begin{equation*}\label{indicator}
\mathbb{I}_{A}(z) := \begin{cases}
1 & \text{ if } z \in A,\\
0 & \text{ otherwise. } 
\end{cases}
\end{equation*}
The ePMF can be used to estimate $\psi$, which gives us $\hat{\psi}=\phi\left( \hat{\theta} \right) = \phi\left( \hat{\mathbb{P}}_X \right)$. This is called a ``plug-in'' estimator, as we plug the estimator $\hat{\theta}$ of $\theta$ (\textit{i.e.} $\hat{\mathbb{P}}_X$ of $\mathbb{P}_X$) into the function $\phi$. In the above example, this implies calculating:

\begin{eqnarray*}
 \hat{\psi} = \phi \left( \hat{\theta} \right) = \phi \left( \hat{\mathbb{P}}_X \right)  = \sum_{x \in \mathcal{X}} x \cdot \hat{\mathbb{P}
}_X (X = x)
\end{eqnarray*}
which, for an observed dataset $\{x_1, x_2, \dots, x_n\}$ is equivalent to taking its mean\cite{Vaart1998}:
\begin{eqnarray*}
 \hat{\psi} = \sum_{x \in \mathcal{X}} x \cdot \hat{\mathbb{P}
}_X (X = x) = \sum_{x \in \mathcal{X}} x \Bigg[ \frac{1}{n} \sum\limits_{i = 1}^n \mathbb{I}_{\{x_i\}}(x)\Bigg] = \frac{1}{n} \sum\limits_{i = 1}^n \sum_{x \in \mathcal{X}} x \cdot \mathbb{I}_{\{x_i \}}(x) = \frac{1}{n} \sum\limits_{i = 1}^n x_i
\end{eqnarray*}
where the last equality follows from the fact that $\mathbb{I}_{\{x_i \}}(x) = 1$ only when $x_i = x$ and in that case the product $x \cdot \mathbb{I}_{\{x_i \}}(x)$ is $x_i \cdot 1 = x_i$ (we exchange $x$ with $x_i$ by using that $x_i = x$ in this scenario). The cases where $x \neq x_i$ don't appear in the sum as $\mathbb{I}_{\{x_i \}}(x) = 0$ results in adding $0$ to the sum.

The functional notation of $\psi$ allows us to study the robustness of our estimations using Hadamard derivatives. In particular, if the data are distributed according to the mass $\mathbb{P}_X$ we can study the rate of change from distribution $\mathbb{P}_X$ in the direction of another distribution, $Q$, by  analyzing the derivative:
\begin{equation*}
 \partial_{Q - \mathbb{P}_X} \phi(\mathbb{P}_X) = \lim_{\substack{h \downarrow 0\\h \neq 0}} \dfrac{\phi\big((1 - h) \mathbb{P}_X + h \cdot Q \big) - \phi(\mathbb{P}_X)}{h}
\end{equation*}
where we have substituted $G_h = Q$ for all $h$ and $\theta = \mathbb{P}_X$ in \eqref{hadamardos}. 

Intuitively this quantifies the rate of change in $\phi$ if the model deviates a little from $\mathbb{P}_X$ towards $Q$ (for example in the case of noisy data). Choosing $Q = \mathbb{I}_{\{Y\}}$ as the indicator of the set that only contains the value $Y$ \eqref{indicator} we can study the rate of change of $\phi$ in the direction of an observation, $Y$. In particular $\mathbb{I}_{\{Y\}}$ stands for the model that assigns probability $1$ to $X$ taking the value $Y$.  Hence the derivative analyzes how an observation, $Y$, \textit{influences} our estimation of $\psi$. 

The Hadamard derivative, in this special case, is called the \textbf{influence function (IF) of the functional $\phi$ under model $\mathbb{P}_X$ at $Y$} and is denoted:
\begin{equation}
 \text{IF}_{\phi,\mathbb{P}_X}(Y) := \partial_{\mathbb{I}_{\{Y\}}- \mathbb{P}_X} \phi(\mathbb{P}_X) = \lim_{\substack{h \downarrow 0\\h \neq 0}} \dfrac{\phi\big((1 - h)\cdot \mathbb{P}_X + h \cdot \mathbb{I}_{\{Y\}} \big) - \phi(\mathbb{P}_X)}{h}.\label{IF}
\end{equation}
\\

The IF stands for the Hadamard derivative in a special case, thus the Taylor expansion in \eqref{taylorhadamard} can be rewritten as:

\begin{equation}
\underbrace{\phi(\hat{\mathbb{P}}_X)}_{\hat{\psi}} \approx \underbrace{\phi(\mathbb{P}_X)}_{\psi} + \underbrace{\text{IF}_{\phi,\mathbb{P}_X}(Y)}_{\partial\phi_{\mathbb{I}_{\{Y\}} - \mathbb{P}_X}(\mathbb{P}_X)}.
\end{equation}

Note that the Hadamard derivative establishes the change of value of a parameter $\psi = \phi(\theta)$ (written as a functional) resultant from small perturbations of the estimator in the direction of $Y$ . Plotting the IF provides a tool to discover outliers and is informative about the robustness of the estimator $\hat{\psi}_n = \psi(\hat{\theta}_n)$.
Finally, if the difference $\hat{\theta}_n - \theta$ is (asymptotically) normally distributed, the Delta-method implies that:
\begin{equation}\label{deltamethodclt}
\hat{\psi}_n - \psi = \phi(\hat{\theta}_n) - \phi(\theta) \overset{\text{approx}}{\sim} \text{Normal}\bigg(0,  \textrm{Var}\Big[\text{IF}_{\phi,\mathbb{P}_X}(Y)\Big]\bigg)
\end{equation}
where the variance, $\textrm{Var}[\text{IF}_{\phi,\mathbb{P}_X}(Y)]$, is taken with respect to the random variable $Y$ (with mass $\mathbb{P}_X$). We remind the reader that an estimator for such a variance given by a random sample $\{X_1, \dots, X_n\}$ is:
\begin{equation}\label{varsimplified}
\widehat{\textrm{Var}}[\text{IF}_{\phi,\mathbb{P}_X}(Y)] = \frac{1}{n}\sum\limits_{i = 1}^n \big( \text{IF}_{\phi,\mathbb{P}_X}(X_i)\big)^2.
\end{equation}
Notice that this estimator is the classical variance estimator $S^2$ for when the mean is known (the mean of the influence function is always $0$).

\subsection{Summary}

The Delta-method to estimate the SE of any particular estimator $\hat{\psi}$ of $\psi$ -- a Hadamard-differentiable function $\psi := \phi(\theta)$ of a parameter $\theta$ -- can be summarized in the following steps:

\begin{enumerate}
    \item\label{step1} Determine the asymptotic distribution of $v := r_n ( \hat{\theta}_n - \theta)$. This variable, $v$, is a function of the distance between the estimator $\hat{\theta}_n$ and the true value $\theta$.
    \item\label{step2} Define the function  $\phi$ related to the scientific question of interest, and compute its Hadamard derivative. Usually $\phi'$ can be obtained from the mass $\mathbb{P}_X$ or the distribution $F_X$ (i.e. the CDF). Recall that in the case of real valued functions $\partial_v \phi(\theta)$ coincides with the classical derivative in the direction of $v$ as in equation \eqref{hadamard}.
    \item\label{step3} Use the asymptotic distribution of $v = r_n(\hat{\theta}_n - \theta)$ obtained in step \ref{step1} and multiply it by the Hadamard derivative in step two. Then, estimate the variance of the distribution and compute the confidence intervals accordingly. Note that in most cases (\textit{e.g.} when  $\phi$ comes from $\mathbb{P}_X$), the difference $r_n(\hat{\theta}_n - \theta)$ is approximately normal and Wald-type confidence intervals can be constructed using the variance in \eqref{varsimplified}, i.e. by estimating the variance through the sample variance of the estimated IF to derive the SE of $\phi(\hat{\theta})$\citep{Agresti2012ApproximateProportions}.
\end{enumerate}

\section{Examples}
In the following sections we'll provide several examples and R code in a set of 6 boxes of applications of the classical and functional Delta-method based on the Hadamard derivative and the IF. The code in the boxes can be accessed at \hyperlink{https://github.com/migariane/DeltaMethodInfluenceFunction}{https://github.com/migariane/DeltaMethodInfluenceFunction}. All calculations and analytical derivations for the classical method were verified using the \texttt{sympy} package\cite{meurer2017sympy} in Python 3.7 in a notebook\cite{python} which can be accessed either in the same repository or in our Google Collab: \hyperlink{https://github.com/migariane/DeltaMethodInfluenceFunction/tree/main/CalculationsDerivationsSympy}{https://github.com/migariane/DeltaMethodInfluenceFunction/tree/main/CalculationsDerivationsSympy}. 

\subsection{Derivation of the Standard Error for the Sample Mean based on the Influence Function (Classical Delta-method)}\label{samplemean}
In this section we derive the standard error for the sample mean.
We illustrate how to apply the proposed steps practically, i.e. by applying equations \eqref{hadamard}, \eqref{jacobian} and \eqref{approxTaylorv1}. Note that the classical statistical inference for the sample mean is straightforward, but the interest here is to show how to derive the IF for the sample mean to then compute the SE applying the steps highlighted before. To derive the SE of the mean for a random sample $\{X_1, X_2, \dots, X_n\}$ we proceed as follows:  First (Step \ref{step1}), we find the distribution of the difference between the estimator $\hat{\theta}_n = \frac{1}{n}\sum\limits_{i = 1}^n X_i$ and the parameter $\theta = \mu$. We know from the central limit theorem that 
$$
\sqrt{\frac{n}{\sigma^2}} \cdot \big( \hat{\theta}_n - \theta\big) \overset{\text{approx}}{\sim}\text{Normal}(0, 1).
$$

In this case, $\phi$ corresponds to the identity function:
$\phi(\hat{\theta}_n)=\hat{\theta}_n$. Then, following Step \ref{step2}, we calculate the Hadamard derivative which in this case corresponds to the classical derivative $\frac{\partial \phi}{\partial\theta}$ in the direction of $\hat{\theta} - \theta$. Hence, following (\ref{eqn:hadamard_from_frechet}), we have:
$$
\partial_{\hat{\theta} - \theta} \phi(\theta) = \frac{\partial \phi}{\partial\theta} \cdot (\hat{\theta} - \theta) = 1 \cdot (\hat{\theta} - \theta)
$$

\noindent We use Taylor's expansion around $\phi(\theta)$ to obtain:

\begin{equation}
\phi(\hat{\theta}) = \phi(\theta)  + \partial_{\hat{\theta} - \theta} \phi(\theta) =  \phi(\theta) + \underbrace{\frac{\partial \phi}{\partial\theta} \left(\frac{1}{n}\sum_{i=1}^{n} X_{i}-\theta\right)}_{\text{IF}_{\phi,\theta}(X)} =
\theta + 1 \cdot \left(\frac{1}{n}\sum_{i=1}^{n} X_{i}-\theta\right) = \frac{1}{n}\sum_{i = 1}^n X_i.
\end{equation}

\noindent Due to the asymptotic normality we can use \eqref{deltamethodclt} to proceed with Step 3:
$$
\phi(\hat{\theta}) - \phi(\theta) \approx \text{Normal}\big(0, \text{Var}[\text{IF}_{\phi,\theta}(X)]\big)
$$
The variance of the influence function is
\begin{equation}
\text{Var}[\text{IF}_{\phi,\theta}(X)] = \text{Var}\left[\frac{1}{n}\sum_{i = 1}^n X_i \right] =\frac{1}{n^2} \sum_{i = 1}^n \text{Var}\left[ X_i \right] = \frac{\sigma^2}{n}
\end{equation}
and thus:
\begin{equation}\label{meanclassic}
    \phi(\hat{\theta}) - \phi(\theta) \approx \text{Normal}\big(0, \frac{\sigma^2}{n} \big)
\end{equation}

The variance of the influence function can be estimated via $\widehat{\text{Var}}[IF_{\phi,\theta}(X)]$ using the standard estimator of the variance, i.e. $S^2 = \frac{1}{n-1}\sum\limits_{i = 1}^n (X_i - \hat{\theta})^2$: 
\begin{equation}
\widehat{\text{Var}}[IF_{\phi,\theta}(X)] = \frac{S^2}{n}
\end{equation}
Two-sided confidence intervals for $\mu$ can thus be estimated through
$$
\hat{\theta} \pm Z_{1 - \alpha/2}\sqrt{\frac{S^2}{n}}
$$
This shows how to obtain the results which are widely known from textbooks through the use of the IF.
\\

\noindent In Box 1 we provide the code to compute the SE for a sample mean using the IF and compare the results with the Delta-method implementation from the R package MSM \citep{kavroudakis2015} and in Figure 1 we plot the IF for the sample mean.
\\

\noindent \textbf{Box 1}. Derivation of the IF for the sample mean
\begin{lstlisting}
# Data generation
set.seed(7777)
n <- 1000
y <- runif(n, 0, 1) 
theoretical_mu <- 0.5 # (1 - 0) / 2 = mu 
empirical_mu   <- mean(y)
# Functional delta-method: estimated influence function for the sample mean (first derivative=1(constant))
IF <- 1 * (y - empirical_mu)
mean(IF) #zero by definition
# Plug-in estimation of the sample mean
Yhat <- y + IF # Plug-in estimator
mean(Yhat)
# Geometry of the IF
plot(y, IF)
# Standard Error: Influence Function
varYhat.IF <- var(IF) / n 
seIF <- sqrt(varYhat.IF);seIF
# 0.009161893

# Asymptotic linear inference 95% Confidence Intervals
Yhat_95CI <- c(mean(Yhat) - qnorm(0.975) * sqrt(varYhat.IF), mean(Yhat) + qnorm(0.975) * sqrt(varYhat.IF)); 
mean(Yhat)
## 0.508518
Yhat_95CI
## 0.490561 0.526475

# Compare with implemented delta-method in library msm 
library(msm)
se <- deltamethod(g = ~ x1, mean = empirical_mu, cov = varYhat.IF)
se
## 0.009161893

# Compare 95%CI delta-method computed by hand with delta-method implemented in RcmdrMisc library
library(RcmdrMisc)
DeltaMethod(lm(y ~ 1), "b0")
#       Estimate SE           2.5%   97.5%
#    b0 0.508518 0.009161893 0.490561 0.526475
\end{lstlisting}

\begin{figure}[ht!]
\begin{center}
\caption{\normalsize. Influence function for the sample mean from the example of Box 1. The IF for the sample mean is unbounded with mean zero. It reflects the deviation of every single sample observation from the empirical mean value i.e., representing the robustness of the estimator for the sample mean against outliers. Furthermore, the graphical display of the IF’s for different estimators will give a clear visual picture of the differences in the data sensitivity of the various estimators.}{\includegraphics[scale=0.5]{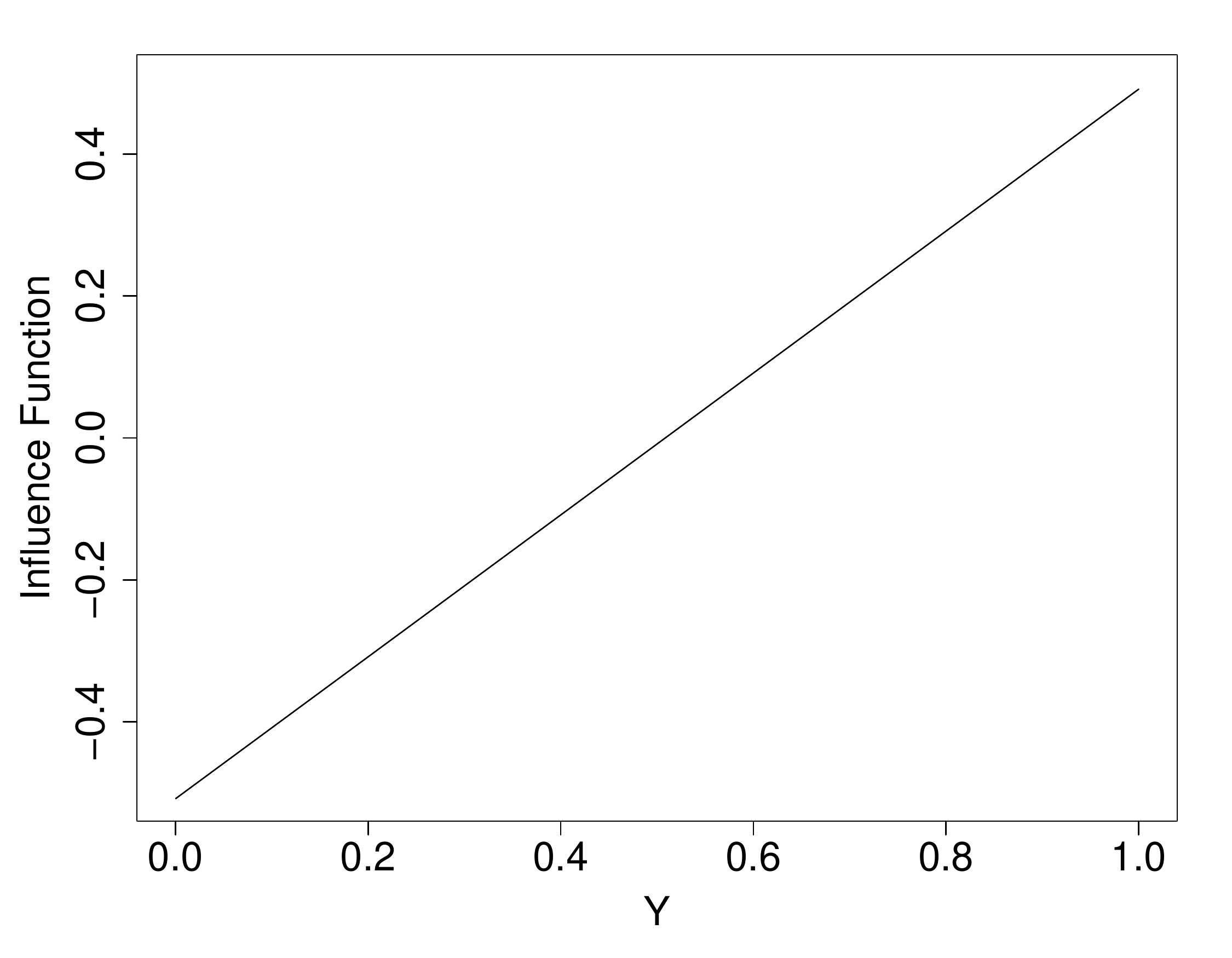}\hfill}
\end{center}
\end{figure}

\subsection{Derivation of the Standard Error for the Sample Mean seen as a Functional (Functional Delta-Method) based on the Influence Function}\label{samplemeandm}

To develop the intuition of how to use the functional delta method we first derive the IF for the sample mean as in section 3.1 but writing the mean $\bar{X}$ as a functional. Afterwards we'll derive the IF for a more complicated situation: the quantile function. 

Consider again the problem of estimating the mean. From the empirical probability mass function, $\mathbb{P}_n$ we obtain the empirical mean, $\bar{X}$, as a functional of $\mathbb{P}_n$. Here we are considering $\phi(\mathbb{P}_n) := \bar{X}$. To simplify the example, assume that the $X_i$ are sampled from a discrete probability mass function $\mathbb{P}$ such that there are only $N$ possible values of the $X_i$. In this case following step \ref{step1} we know from the central limit theorem that for each value $x$, the difference between the empirical probability mass function (which is an average) and its true value is asymptotically normal:
\begin{equation}\label{asymptdist}
  \lim_{n \to \infty} \sqrt{n}\big(\mathbb{P}_n - \mathbb{P}\big) \sim \text{Normal}\big(0,\mathbb{P} \cdot ( 1- \mathbb{P})\big).  
\end{equation}

Where we have defined the empirical probability mass function as in \eqref{epmf}:
$$
\mathbb{P}_n(x) =
\dfrac{\text{Number of times }x \text{ is in the sample}}{n} =\frac{1}{n} \sum\limits_{k = 1}^n \mathbb{I}_{\{x_i\}}(x),
$$
where the indicator variables are defined in section \ref{if}. We remark that the variance from \eqref{asymptdist} results from the variance of the indicators $\mathbb{I}_{\{X_i\}}(x)$ which are Bernoulli distributed. 

We then follow step \ref{step2} to write the functional $\phi$ in terms of the estimator. In this case, the population mean is written as:
$$
\phi(\mathbb{P}):= \mu = \sum\limits_{i = 1}^N x_i \mathbb{P}(x_i).
$$
while the sample mean is given by the following expression:
$$
\phi(\mathbb{P}_n) =  \hat{\mu} = \sum\limits_{i = 1}^N x_i \mathbb{P}_n(x_i)
$$
We remark that in this case we will use the functional delta method as $\phi$ is a functional of the function $\mathbb{P}$. Hence to obtain the approximation in this case (step \ref{step3}) we calculate the influence function from the definition in \eqref{IF}:
\begin{equation}
\begin{aligned}
\textrm{IF}_{\phi,\mathbb{P}_X}(Y) & = \lim_{\substack{h \downarrow 0\\h \neq 0}} \dfrac{\phi\big((1 - h)\cdot \mathbb{P}_X + h \cdot \mathbb{I}_{\{Y\}} \big) - \phi(\mathbb{P}_X)}{h} 
\\ & = \lim_{\substack{h \downarrow 0\\h \neq 0}}\frac{\sum\limits_{i = 1}^N x_i \big( (1 - h)\cdot \mathbb{P}(x_i) + h \cdot \mathbb{I}_{\{Y\}}(x_i) \big) - \sum\limits_{i = 1}^N x_i \cdot \mathbb{P}(x_i)}{h} 
\\ & = \sum\limits_{i = 1}^N x_i \cdot \mathbb{I}_{\{Y\}}(x_i) - \sum\limits_{i = 1}^N x_i \cdot \mathbb{P}(x_i) 
\\ & = \phi(\mathbb{I}_{\{Y\}}) - \phi(\mathbb{P})
\\ & = Y - \phi(\mathbb{P})
\end{aligned}
\end{equation}
Finally the variance of the influence function corresponds to the variance of $Y$:
\begin{equation}
    \textrm{Var}\big[\textrm{IF}_{\phi,\mathbb{P}_X}(Y)\big] = \textrm{Var}\big[Y\big] = \sigma^2
\end{equation}
hence:
\begin{equation}\label{asymptdist}
  \lim_{n \to \infty} \sqrt{n}\big(\phi(\mathbb{P}_n) - \phi(\mathbb{P})\big) \sim \text{Normal}\big(0,\sigma^2\big).  
\end{equation}
which is equivalent to the expression found by the classical method in \eqref{meanclassic}. 

\subsection{Derivation of the Standard Error for the Ratio of Two Means}

Consider a random sample of size $n$ of the i.i.d random variables $X$ and $Y$, which are both normally distributed, with respective means $\mu_X$ and $\mu_Y$ which are estimated by their sample means $\bar{X}$ and $\bar{Y}$. We are interested in deriving the variance for the ratio of the two means (\textit{i.e.} the ratio estimator) defined as: $\phi(\mu_X, \mu_Y) = \frac{{\mu}_{X}}{{\mu}_{Y}}$. In this case (following step \ref{step1}) it is known that the difference $ \hat{\theta} - \theta $ is asymptotically normal. 

Second (step \ref{step2}) we obtain the Hadamard derivative which in this case corresponds to the gradient in the direction of
\begin{equation*}
    v = \hat{\theta} - \theta = \begin{pmatrix}
    \bar{X}\\
    \bar{Y}
    \end{pmatrix} - \begin{pmatrix}
    \mu_X\\
    \mu_Y
    \end{pmatrix} = 
    \begin{pmatrix}
    \bar{X} - \mu_X\\
    \bar{Y} - \mu_Y
    \end{pmatrix}.
\end{equation*}
The gradient is given by:
\begin{equation*}
\nabla \phi = 
\begin{pmatrix}
  \frac{\partial \phi}{\partial\mu_{X}} \\[1em]
  \frac{\partial \phi}{\partial\mu_{Y}} \\
\end{pmatrix}
= 
\begin{pmatrix}
  \frac{1}{\mu_{Y}} \\[1em]
  - \frac{\mu_{X}}{\mu_{Y}^{2}}
\end{pmatrix}
\end{equation*}
where we assume $\mu_{Y} \neq 0$. The Hadamard derivative (\textit{i.e.} the influence function) is given by:
\begin{equation*}
\text{IF}_{\phi,P}(X,Y) = \partial_v \phi(\bar{X},\bar{Y}) = \Big( \frac{1}{\mu_{Y}},  - \frac{\mu_{X}}{\mu_{Y}^{2}}\Big)  \begin{pmatrix}
    \bar{X} - \mu_X\\
    \bar{Y} - \mu_Y
    \end{pmatrix} = \frac{1}{\mu_Y}(\bar{X} - \mu_X) -  \frac{\mu_X}{\mu_Y^2}(\bar{Y} - \mu_Y).
\end{equation*}
The variance is hence given by the variance of the influence function (\textit{i.e.} the Hadamard derivative): 
\begin{equation}
\begin{aligned}
\textrm{Var}\Big(\text{IF}_{\phi,P}(X,Y)\Big) & = \textrm{Var}\Big( \frac{1}{\mu_Y}(\bar{X} - \mu_X) - \frac{\mu_X}{\mu_Y^2}(\bar{Y} - \mu_Y)\Big) = \frac{1}{n}\bigg( \frac{1}{\mu_Y^2} \textrm{Var}(X) + \frac{\mu_X^2}{\mu_Y^4} \textrm{Var}(Y) - 2 \frac{\mu_X}{\mu_Y^3} \textrm{Cov}(X,Y) \bigg)
\end{aligned}    
\end{equation}
where we used that $\text{Var}(\bar{X}) = \text{Var}(\frac{1}{n} \sum_i X_i) = \frac{1}{n^2} \cdot \text{Var}(X_1+ \ldots +X_n) = \frac{1}{n} \cdot$ Var(X) under the independence assumption, $\text{Var}(bX)=b^2\,$ Var(X) and  $\text{Var}(\mu_{\cdot}) = 0$. 

For step \ref{step3}, the estimated standard error is then obtained as the square root of the estimated variance and Wald-type confidence intervals (level $(1 - \alpha)\times 100\%$) follow:
\begin{equation*}
 \frac{\bar{X}}{\bar{Y}} \pm Z_{1  - \alpha/2} \sqrt{\widehat{\textrm{Var}}\Big(\text{IF}_{\phi,P}(X,Y)}\Big),
\end{equation*}
where the estimator for the variance is:

\begin{equation*}
\widehat{\textrm{Var}}\Big(\text{IF}_{\phi,P}(X,Y)\Big) = \frac{1}{n}\bigg( \frac{1}{\bar{Y}^2} \widehat{\textrm{Var}}(X) + \frac{\bar{X}^2}{\bar{Y}^4} \widehat{\textrm{Var}}(Y) - 2 \frac{\bar{X}}{\bar{Y}^3} \widehat{\textrm{Cov}}(X,Y) \bigg).
\end{equation*}
\\

\noindent \textbf{Box 2}. Derivation of the IF for the ratio of two sample means
\begin{lstlisting}
# Data generation
library(mvtnorm)
set.seed(123)
data  <- as.data.frame(rmvnorm(1000, c(3,4), matrix(c(1,0.3,0.3,2), ncol = 2)))
colnames(data) <- c("X","Y")
data <- as.data.frame(data)
attach(data)
# SE estimation for the ratio: Delta-Method based on the IF
ratio <- mean(X)/mean(Y);mean(ratio)
[1] 0.7541776
n <- 1000
a <-    (1 / (mean(Y))^2) * var(X) 
b <-   ((mean(X))^2 / (mean(Y))^4) * var(Y) 
c <-    2 * ((mean(X)) / (mean(Y))^3) * cov(X,Y)
var.IF <- 1/n *(a+b-c); var.IF
[1] 0.000111181
SE <- sqrt(var.IF); SE
[1] 0.01054424
CI = c(mean(ratio)-qnorm(0.975)*SE,mean(ratio)+qnorm(0.975)*SE); mean(ratio); CI
[1] 0.7541776
[1] 0.7335113 0.7748440
# Cheking results
# CI Delta method
theta1 <- mean(X)
sd1 <- sd(X)
theta2 <- mean(Y)
sd2 <- sd(Y)
CI.Delta = function(theta1, sd1,
                     theta2, sd2,# estimate and estimated sd of estimator
                     alpha # theoretical coverage (1-alpha)
 )
 {
## CI Delta method
## Hirschberg, J., and J. Lye. 2010.
## “A Geometric Comparison of the Delta and Fieller Confidence Intervals.”
## The American Statistician 64 (3): 234–41
## We assume here independence between theta1.hat and theta2.hat
    z = qnorm(p=1-alpha/2)
    R = theta1/theta2
    sd.R = sqrt(sd1^2 + R^2 * sd2^2)/theta2
    L = R - z*sd.R
    U = R + z*sd.R
    res = c(R,L,U)
    names(res) = c('R','L','U')
    return(res)
 }
# 95%CI Delta-method
CI.Delta(theta1, sd1, theta2, sd2, 0.95)
        R         L         U 
0.7541776 0.7315508 0.7768045 
\end{lstlisting}

\subsection{Derivation of the Standard Error for the Ratio of Two Probabilities (Risk Ratio)}\label{samplemean}

In medical statistics, we are often interested in marginal and conditional (sometimes causal) risk ratios. Consider Table 1, where we are interested in the mortality risk by cancer status. Let $p_1$ denote the probability of being alive given that the patient has cancer and $p_2$ the probability of being dead given that the patient suffers the disease. 

\begin{longtable}[]{@{}rrr@{}}
\toprule
\begin{minipage}[b]{0.24\columnwidth}\raggedright
Risk\strut
\end{minipage} & \begin{minipage}[b]{0.33\columnwidth}\raggedright
Alive\strut
\end{minipage} & \begin{minipage}[b]{0.33\columnwidth}\raggedright
Dead\strut
\end{minipage}\tabularnewline
\midrule
\endhead
\begin{minipage}[t]{0.24\columnwidth}\raggedright
Cancer\strut
\end{minipage} & \begin{minipage}[t]{0.33\columnwidth}\raggedright
$p_{1}$  \strut
\end{minipage} & \begin{minipage}[t]{0.33\columnwidth}\raggedright
$p_{2}$   \strut
\end{minipage}\tabularnewline
\begin{minipage}[t]{0.24\columnwidth}\raggedright
No cancer\strut
\end{minipage} & \begin{minipage}[t]{0.33\columnwidth}\raggedright
$1 - p_1$  \strut
\end{minipage} & \begin{minipage}[t]{0.33\columnwidth}\raggedright
$1 - p_2$   \strut
\end{minipage}\tabularnewline
\begin{minipage}[t]{0.24\columnwidth}\raggedright
Size of sample\strut
\end{minipage} & \begin{minipage}[t]{0.33\columnwidth}\raggedright
\(\text{n}_{1}\)\strut
\end{minipage} & \begin{minipage}[t]{0.33\columnwidth}\raggedright
\(\text{n}_{2}\)\strut
\end{minipage}\tabularnewline
\bottomrule
\caption{\label{tab:Table 1} Classical 2x2 epidemiological cross table}
\end{longtable}

Suppose we are interested in estimating a two-sided confidence interval for the cancer mortality risk ratio (RR), i.e.:
\begin{equation*}
\text{RR}(p_1, p_2) = \frac{p_1}{p_2}\,.
\end{equation*}
Thus, $\phi$ relates to the ratio of the two probabilities. To now derive the risk ratio's confidence interval through the proposed steps, it is better to work with the natural logarithm of the risk ratio:\citep{Selvin2009EpidemiologicApproach}
\begin{equation*}
\phi(p_1, p_2) = \ln\big(\text{RR}(p_1, p_2)\big) =  \ln\left(\frac{p_1}{p_2}\right)= \ln(p_{1}) - \ln( p_{2})\,.
\end{equation*}
The steps are the same as before: we compute the Hadamard derivative (gradient) in the direction of 
\begin{equation*}
    v = \hat{p} - p = 
    \begin{pmatrix}
    \hat{p_1} - p_1\\
    \hat{p_2} - p_2\\
    \end{pmatrix}
\end{equation*}

where we know that $v$ is approximately normally distributed if $p$ is not extremely low or high, and  if the condition
$np(1-p)\geq 9$ is fulfilled  (step \ref{step1}).\citep{RN2073}

For step \ref{step2}, the Hadamard (directional) derivative is:
\begin{equation*}
\text{IF}_{\phi,P}(X,Y) = \nabla \phi(p_1,p_2)^T v = 
\left(\frac{1}{p_1},-\frac{1}{p_2}\right) \begin{pmatrix}
    \hat{p_1} - p_1\\
    \hat{p_2} - p_2\\
    \end{pmatrix}
=  \frac{\hat p_1}{p_1}  - \frac{\hat p_2}{p_2} \,.
\end{equation*}

Finally, under the assumption that $\hat{p_1}$ and $\hat{p_2}$ are independent (i.e. Cov($\hat{p_1}$, $\hat{p_2}$)$=0$), using that $\text{Var}(\hat{p_i}) = p_i (1 - p_i)/n_i$, $\text{Var}(bX)=b^2\, \text{Var}(X) \, \text{and} \, \text{Var}(p_i)=0$, its variance is given by:
\begin{equation}
    \textrm{Var}\underbrace{\Big[\nabla \phi(p_1,p_2)^T v\Big]}_{\text{IF}_{\phi,P}(X,Y)} = \frac{1}{p_{1}^{2}}\frac{ p_{1}(1- p_{1})}{n_{1}}+ \frac{1}{ p_{2}^{2}}\frac{p_{2}(1- p_{2})}{n_{2}}\,.
\end{equation}

An estimator for the variance in the confidence intervals (for step \ref{step3}) is then:
\begin{equation}
    \widehat{\textrm{Var}}\Big[\nabla \phi(p_1,p_2)^T v\Big] = \frac{1}{\hat{p}_{1}^{2}}\frac{ \hat{p}_{1}(1- \hat{p}_{1})}{n_{1}}+ \frac{1}{ \hat{p}_{2}^{2}}\frac{\hat{p}_{2}(1- \hat{p}_{2})}{n_{2}}.\\
\end{equation}
Hence the Wald-type confidence interval is given by:
$$
\ln\big(\text{RR}(\hat{p}_1, \hat{p}_2)\big) \pm Z_{1  - \alpha/2} \sqrt{\frac{1}{\hat{p}_{1}^{2}}\frac{ \hat{p}_{1}(1- \hat{p}_{1})}{n_{1}}+ \frac{1}{ \hat{p}_{2}^{2}}\frac{\hat{p}_{2}(1- \hat{p}_{2})}{n_{2}}}.
$$
If we then exponentiate the confidence limits, we get the confidence interval for the risk ratio.\\

\noindent \textbf{Box 3}. Derivation of the IF for the ratio of two independent proportions
\begin{lstlisting}
#Data generation
install.packages("epitools")
library(epitools)
RRtable <- matrix(c(60,40,40,60),nrow = 2, ncol = 2)
RRtable
# The next line asks R to compute the RR and 95% confidence interval
riskratio.wald(RRtable)

#$data
#          Outcome
#Predictor  Disease1 Disease2 Total
#  Exposed1       60       40   100
#  Exposed2       40       60   100
#  Total         100      100   200
#
#$measure
#          risk ratio with 95% C.I.
#Predictor  estimate    lower    upper
#  Exposed1      1.0       NA       NA
#  Exposed2      1.5 1.124081 2.001634

# SE estimation for the ratio
p1 <- 0.6
p2 <- 0.4
N1 <- 100
N2 <- 100
ratio <- 0.6 / 0.4; ratio
# 1.5

var.IF <- (1 / (p1)^2 * (p1 * (1 - p1)/ N1)) + (1 / (p2)^2 * (p2 * (1 - p2)/ N2))
# 0.02166667
SE <- sqrt(var.IF); SE
# 0.147196
CI = c(log(ratio)-qnorm(0.975)*SE,log(ratio)+qnorm(0.975)*SE); ratio; exp(CI)
# 1.5
# 1.124081 2.001634
\end{lstlisting}

\subsection{Counterexample: When the assumptions for Using the Influence Function do not hold}\label{counterexample}
The Delta-method will not work for non-differentiable functions, such as indicators, maximums, absolute values, and others\citep{Kennedy2022SemiparametricReview}. However, there are some functions for which, even if differentiable, it is not guaranteed that the Taylor Series expansion and thus the functional Delta-method are the ideal approach to derive the SE. For such differentiable functions, the functional Delta-method can provide incorrect results. Consider the attributable fraction (AF) among the exposed for (a continuous) exposure level $x$ given by:
$$
\phi(\theta) := \textrm{AF}_e(x;\theta) = \dfrac{RR(x; \theta)-1}{RR(x; \theta)}
$$
where $RR(x; \theta)$, refers to the (continuous) risk ratio between different levels of exposure $x$. The value $\theta$ is a parameter.  
\\

Let the relative risk be defined as:
$$
RR(x; \theta) = e^{x/\theta} \qquad \textrm{for } x > 0
$$
where as the exposure level $x$ increases, the risk grows as well (as in the relationship between second-hand smoke inhalation and lung cancer).
The AF among the exposed is given by:
$$
\textrm{AF}_e(x;\theta) =
\begin{cases}
\dfrac{e^{x/\theta}-1}{e^{x/\theta}} & \textrm{ if } \theta > 0, \\
1 & \textrm{ if } \theta \leq 0.
\end{cases}
$$

For this type of modeling it is common to use a previously estimated parameter $\theta$ (say from a previous study) which usually comes from  maximum likelihood estimation (MLE). We know from MLE theory that $\hat\theta$ is asymptotically normal with the variance given by Fisher's information coefficient. Hence we have completed step \ref{step1}.

For step \ref{step2} we notice that the function $\phi(\theta) = \textrm{AF}_e(x;\theta)$ is continuous and differentiable. The left side limit $\lim_{\theta\to0^-}\phi(\theta)  = 1$ as $\theta \leq 0$ in that case and the right hand side limit is $\lim_{\theta\to0^+}\phi(\theta)  = \lim_{\theta\to0^+} 1 - e^{-x/\theta}= 1 - e^{-\infty} = 1$ where the penultimate equality happens because $\lim_{\theta\to0^+} 1/\theta=\infty$. To check differentiability at $0$ we check that the left and right limits coincide: $\phi'_+(0) = \lim_{h \downarrow 0} \frac{\phi(0 + h) - \phi(0)}{h} = \lim_{h \downarrow 0} \frac{1 - e^{-x/h} - 1}{h} = \lim_{h \downarrow 0} -  \frac{e^{-x/h}}{h} =  \lim_{h \downarrow 0} - \frac{1}{h} e^{-1/h \cdot x} = 0$ as the exponential converges faster than the fraction to $e^{-\infty} = 0$. On the other hand, the left limit is: $\phi'_-(0) = \lim_{h \downarrow 0} \frac{\phi(0 - h) - \phi(0)}{h} = \lim_{h \downarrow 0} \frac{1 - 1}{h} = 0$. As both limits coincide, the derivative exists and is zero. Henceforth, we can obtain its derivative:
$$
\phi'(\theta) = \dfrac{\partial \textrm{AF}_e}{\partial\theta} = \begin{cases}
\dfrac{x}{\theta^2}\cdot \Bigg( \dfrac{e^{x/\theta}-1}{e^{x/\theta} \cdot \theta^2}  - 1\Bigg)& \textrm{ if } \theta > 0 \\
0 & \textrm{ if} \theta \leq 0.
\end{cases}
$$

The derivative works well in cases where $\theta > 0$. However in extreme cases where all (or almost all) of the incidents in the exposed group can be attributed the exposure we have $\textrm{AF}_e \approx 1$ which in turn implies either $\theta \leq 0$ or $\theta \approx 0$ and if Taylor's expansion (step \ref{step3}) is done around $\theta = 0$ we obtain:
$$
IF_{\phi,P}(X) =  \dfrac{d \textrm{AF}_e}{dx} \cdot (\hat\theta_n - 0) = 0
$$
which is a terrible approximation if $\theta\approx 0$ but not exactly $0$. 

Therefore, for some continuous and differentiable functions, such as the AF, there are particular situations where the Hadamard derivative at some points might not provide a good approximation for the SE using the IF. 

\subsection{Derivation of the Standard Error for the Quantile Function (Functional Delta-Method) based on the Influence Function }\label{quantile}

 For a continuous random variable with density $f$ and cumulative distribution $F$, it is possible to write a quantile in terms of the inverse of the cumulative distribution:
$$
\phi(F) := F^{-1}(p) = \min\{ x | F(x) \geq p\}
$$
where
$$
\lim_{n\to\infty} \sqrt{n}\big(F_n - F\big) \sim \textrm{Normal}\big(0, F \cdot (1 - F)\big)
$$
thus finding the limiting distribution of step \ref{step1}. 

To obtain the derivative (step \ref{step2}) we use the inverse function relationship between quantiles and cumulative probabilities. Given a quantile $q = F^{-1}(p)$, the probability associated with the quantile can be written as $p = F(q) = F\big(F^{-1}(p)\big)$. 

Let $G_h = (1 - h) \cdot F + h \cdot  \mathbb{I}_{[Y, \infty)}$. Notice that in this case (in contrast with section \ref{samplemeandm}) we use the cumulative distribution of a point pass $\mathbb{I}_{[Y, \infty)}$ and not just its probability, $\mathbb{I}_{\{Y\}}$, as this derivative is taken in the space of cumulative distributions not in the space of probabilities. 

We'll assume the function is invertible and one can obtain the true probability $p$ by the composition:
\begin{equation}\label{pq}
p  = G_h\big( G_h^{-1}(p)\big).
\end{equation}
The chain rule (which also applies for Hadamard derivatives \cite{VanderLaan2011}) allows us to compute the derivative implicitly via the same trick used in elementary calculus courses to find the inverses of the trigonometric and exponential functions (\textit{i.e.} by using the inverse \eqref{pq}):
\begin{equation}
    \begin{aligned}
    0 & = \frac{d}{dh} G_h \big( G_h^{-1}(p)\big) 
    \\ & = f\big(F^{-1}(p)\big) \cdot \partial_{\mathbb{I}_{[Y, \infty)} - F}  F^{-1}(p) + (\mathbb{I}_{[Y, \infty)} - F) \big(F^{-1}(p)\big)
     \\ & = f\big(F^{-1}(p)\big) \cdot \partial_{\mathbb{I}_{[Y, \infty)} - F}  F^{-1}(p)+ (\mathbb{I}_{[Y, \infty)} - F) \big(q\big)
     \\ & = f\big(F^{-1}(p)\big) \cdot \partial_{\mathbb{I}_{[Y, \infty)} - F}  F^{-1}(p)+ \mathbb{I}_{[Y, \infty)}(q) - F(q)
     \\ & = f\big(q\big) \cdot \partial_{\mathbb{I}_{[Y, \infty)} - F}  F^{-1}(p)+ \mathbb{I}_{[Y, \infty)}(q) - p
    \end{aligned}
\end{equation}
where $\partial_{\mathbb{I}_{[Y, \infty)} - F}$ follows the notation of \ref{hadamardos}. Hence:
$$
\partial_{\mathbb{I}_{[Y, \infty)} - F}  F^{-1}(p) = \frac{\mathbb{I}_{[Y, \infty)}(q) - p}{f\big(q\big)}= \text{IF}(Y) 
$$

The variance is given by:
$$
\textrm{Var}\big[\text{IF}(Y) \big] = 
 \frac{1}{\big(f(q)\big)^2} \textrm{Var}\big[\mathbb{I}_{[Y, \infty)}(q)\big] = \frac{1}{\big(f(q)\big)^2} \mathbb{P}(Y \leq q)\cdot \big(1 - \mathbb{P}(Y \leq q)\big) = \frac{p\cdot (1-p)}{\big(f(q)\big)^2}
$$

As $f$ is unknown, the sample variance of the influence function does not have a closed form; however it can be computed using $\hat{f}$, a kernel density estimate of $f$ as shown in Box 4. \\

\noindent \textbf{Box 4.} Delta-method to derive the SE for the quantile function
\begin{lstlisting}
# Data generation
set.seed(7777)
library(kdensity)
library(EnvStats)
n           <- 1000
y           <- rnorm(n,50)
my_p        <- 0.25 #Change as you see fit

#Compute the first quartile 
empirical_quantile <- quantile(y, my_p); 
f_hat              <- kdensity(y, kernel = "epanechnikov", normalized = F)
plot(f_hat, main = "Estimated density f() of data")

# IF based 95%CI for Y 
var_IF  <- my_p*(1 - my_p)/(f_hat(empirical_quantile)^2)
SEy_IF  <- sqrt(var_IF/n)
CI      <- c(empirical_quantile - qnorm(0.975)*SEy_IF, empirical_quantile + qnorm(0.975)*SEy_IF); CI

## 25%      95%
## 49.28908 49.46078 

# Check results binomial and asymptotically based 95%CI 
eqnpar(x=y, p=my_p, ci=TRUE, ci.method="exact",approx.conf.level=0.95)$interval$limits

##     LCL      UCL 
## 49.26460 49.45987 

eqnpar(x=y, p=my_p, ci=TRUE, ci.method="normal.approx",approx.conf.level=0.95)$interval$limits
##     LCL      UCL 
## 49.26406 49.45863 
\end{lstlisting}

\subsection{Delta-method for the Derivation of the Standard Error of the Correlation Coefficient}

Consider the correlation coefficient $\rho$ between two variables $X$ and $Y$, which is defined as follows (the key to implementing the functional Delta-method here is to represent the correlation parameter $\rho$ in terms of expectations):
\begin{equation}
\rho(X,Y)=\frac{\mathbb{E}[X Y]−\mathbb{E}[X]\cdot\mathbb{E}[Y]}
{\sqrt{\mathbb{E}[X^{2}]-\mathbb{E}^{2}[X]}\sqrt{\mathbb{E}[Y^{2}]-\mathbb{E}^{2}[Y]}}.
\label{RHO}
\end{equation}
Therefore, $\rho$ is a function of different moments of both $X$ and $Y$. For each term in the definition of the correlation coefficient in \eqref{RHO}, we have to estimate: $\mu_{XY} = \mathbb{E}[X Y]$, $\mu_X = \mathbb{E}[X]$, $\mu_Y = \mathbb{E}[Y]$, $\mu_{X^2} = \mathbb{E}[X^2]$, and $\mu_{Y^2} = \mathbb{E}[Y^2]$. Each of those terms can be estimated via their respective plug-in estimators: $\hat{\mu}_{XY} = \overline{XY}$, $\hat{\mu}_X = \overline{X}$, $\hat{\mu}_Y = \overline{Y}$, $\hat{\mu}_{X^2} = \overline{X^2}$, and $\hat{\mu}_{Y^2} = \overline{Y^2}$. We remark that for Step \ref{step1} it is known that the moments are asymptotically normal.\citep{Hayashi2011Econometrics} 

An estimator for $\rho$ is given by (step \ref{step2}):
\begin{equation*}
\hat{\rho}:=\phi(\hat{\mu}_{XY},\hat{\mu}_X,\hat{\mu}_Y,\hat{\mu}_{X^2},\hat{\mu}_{Y^2}) = \dfrac{\hat{\mu}_{XY} - \hat{\mu}_X\cdot\hat{\mu}_Y}{\sqrt{\hat{\mu}_{X^2} - \hat{\mu}_X^2}\cdot \sqrt{\hat{\mu}_{Y^2} - \hat{\mu}_Y^2}}.\label{rho}
\end{equation*}

The influence function (or Hadamard derivative) is computed as the product of the gradient of $\phi$ with the directional vector:
$$
v = \Big( {\hat\mu_{XY}} - \mu_{XY}, \hat{\mu}_X - \mu_X, \hat{\mu}_Y - \mu_Y, \hat{\mu}_{X^2} - \mu_{X^2}, \hat{\mu}_{Y^2} - \mu_{Y^2}\Big)^T.
$$
So, using the principles of equation \eqref{taylorhadamard}, we have to take the partial derivatives to obtain the gradient: 
\begin{equation*}
    \nabla \phi \Big(\mu_{XY}, \mu_X, \mu_Y, \mu_{X^2}, \mu_{Y^2}\Big) = 
    \begin{pmatrix}\frac{1}{\sqrt{- \mu_{X}^{2} + \mu_{X^2}} \sqrt{- \mu_{Y}^{2} + \mu_{Y^2}}}\\\frac{\mu_{X} \left(- \mu_{X} \mu_{Y} + \mu_{XY}\right)}{\left(- \mu_{X}^{2} + \mu_{X^2}\right)^{\frac{3}{2}} \sqrt{- \mu_{Y}^{2} + \mu_{Y^2}}} - \frac{\mu_{Y}}{\sqrt{- \mu_{X}^{2} + \mu_{X^2}} \sqrt{- \mu_{Y}^{2} + \mu_{Y^2}}}\\- \frac{\mu_{X}}{\sqrt{- \mu_{X}^{2} + \mu_{X^2}} \sqrt{- \mu_{Y}^{2} + \mu_{Y^2}}} + \frac{\mu_{Y} \left(- \mu_{X} \mu_{Y} + \mu_{XY}\right)}{\sqrt{- \mu_{X}^{2} + \mu_{X^2}} \left(- \mu_{Y}^{2} + \mu_{Y^2}\right)^{\frac{3}{2}}}\\- \frac{- \mu_{X} \mu_{Y} + \mu_{XY}}{2 \left(- \mu_{X}^{2} + \mu_{X^2}\right)^{\frac{3}{2}} \sqrt{- \mu_{Y}^{2} + \mu_{Y^2}}}\\- \frac{- \mu_{X} \mu_{Y} + \mu_{XY}}{2 \sqrt{- \mu_{X}^{2} + \mu_{X^2}} \left(- \mu_{Y}^{2} + \mu_{Y^2}\right)^{\frac{3}{2}}}\end{pmatrix} = 
    \begin{pmatrix}
    \nabla \phi_1 \\
    \nabla \phi_2 \\
    \nabla \phi_3 \\
    \nabla \phi_4 \\
    \nabla \phi_5 \\
    \end{pmatrix}.
\end{equation*}

The Hadamard derivative (influence function) results in:
\begin{equation*}
\begin{aligned}
\nabla \phi^T v & = ( \overline{XY} - \mu_{XY} )\cdot\nabla\phi_1   + ( \overline{X} - \mu_{X})\cdot\nabla \phi_2 + ( \overline{Y} - \mu_{Y})\cdot \nabla\phi_3 + ( \overline{X^2} - \mu_{X^2})\cdot \nabla\phi_4  + (\overline{Y^2} - \mu_{Y^2})\cdot\nabla\phi_5  \\
& = \frac{1}{n}\sum\limits_{i=1}^n \left(  (X_i Y_i -  \mu_{XY}) \cdot \nabla\phi_1 +  ( X_i - \mu_{X}) \cdot \nabla\phi_2 + ( Y_i - \mu_{Y}) \cdot \nabla\phi_3 +  ( X_i^2 - \mu_{X^2} ) \cdot \nabla\phi_4 + (Y_i^2 - \mu_{Y^2}) \cdot \nabla\phi_5 \right) 
\end{aligned}
\end{equation*}

Hence the variance of the influence function can be computed as: 
\begin{equation*}
\begin{aligned}
\text{Var}\big[ \nabla \phi^T v\big] & = 
\frac{1}{n}\text{Var}\Big[  \underbrace{(X Y -  \mu_{XY}) \cdot \nabla\phi_1 +  ( X - \mu_{X}) \cdot \nabla\phi_2 + ( Y - \mu_{Y}) \cdot \nabla\phi_3 +  ( X^2 - \mu_{X^2} ) \cdot \nabla\phi_4 + (Y^2 - \mu_{Y^2}) \cdot \nabla\phi_5}_{h(X,Y)} \Big] \\
& = 
\frac{1}{n}\text{Var}\left[ h(X,Y) \right]
\end{aligned}
\end{equation*}

An estimator of the variance is then:
\begin{equation*}
\begin{aligned}
\widehat{\text{Var}}\big[ \nabla \phi^T v\big] = 
\frac{1}{n}\widehat{\text{Var}}\left[ h(X,Y) \right]  = \frac{1}{n} \cdot \left[\frac{1}{n-1}\sum\limits_{i=1}^n \big( h(X_i,Y_i) - \overline{h(X,Y)} \big)^2\right]
\end{aligned}
\end{equation*}
where $\overline{h(X,Y)} = \frac{1}{n}\sum\limits_{i = 1}^n h(X_i,Y_i)$ corresponds to the sample mean of the $h$ function. 

Finally, for Step \ref{step3} it is possible to build a $(1 - \alpha)\times 100 \%$ type Wald confidence interval  \citep{Agresti2010} as follows:
\begin{equation*}
\hat \rho \pm Z_{1- \alpha/2}\sqrt{\widehat{\text{Var}}\big[ \nabla \phi^T v\big] }.
\end{equation*}
\\

In Box 5 we show how to compute the SE for $\rho$ using the IF derived manually and  contrast the results with those of the implementation of the Delta-method from the \emph{confintr} R package. A convenient function is called \emph{ci\_cor}, which estimates the type Wald 95\% CI for the correlation parameter. Also, obtaining a CI by bootstrap is supported.
\\

\noindent \textbf{Box 5.} Delta-Method for estimating the SE of the correlation between two variables X and Y using the IF.

\begin{lstlisting}
# Generate the data
#install.packages("MASS")
library('MASS')
samples = 1000
R = 0.83
set.seed(1)
data = mvrnorm(n=samples, mu=c(0, 0), Sigma=matrix(c(1, R, R, 1), nrow=2), empirical=TRUE)
X = data[, 1]  # standard normal (mu=0, sd=1)
Y = data[, 2]  # standard normal (mu=0, sd=1)
# Assess that it works
cor(X, Y)  # r = 0.83

mu1 = mean(X*Y)
mu2 = mean(X)
mu3 = mean(Y)
mu4 = mean(X^2)
mu5 = mean(Y^2) 

IF1 = X*Y-mu1 
IF2 = X-mu2
IF3 = Y-mu3 
IF4 = X^2-mu4 
IF5 = Y^2-mu5

IF = 
    (sqrt(mu4-mu2^2)*sqrt(mu5-mu3^2))^(-1)*IF1+ 
    (-mu3*sqrt(mu4-mu2^2)*sqrt(mu5-mu3^2)+(mu1-mu2*mu3)*mu2*sqrt(mu5-mu3^2)/sqrt(mu4-mu2^2))/((mu4-mu2^2)*(mu5-mu3^2))*IF2+ 
    (-mu2*sqrt(mu4-mu2^2)*sqrt(mu5-mu3^2)+(mu1-mu2*mu3)* mu3*sqrt(mu4-mu2^2)/sqrt(mu5-mu3^2))/((mu4-mu2^2)*(mu5-mu3^2))*IF3+ 
    (-mu1+mu2*mu3)/(2*(mu4-mu2^2)^1.5*(mu5-mu3^2)^.5)*IF4+ 
    (-mu1+mu2*mu3)/(2*(mu4-mu2^2)^.5*(mu5-mu3^2)^1.5)*IF5

SE = sd(IF)/sqrt(1000); SE

rho_hat = (mu1-mu2*mu3)/(sqrt(mu4-mu2^2)*sqrt(mu5-mu3^2)); rho_hat
CI = c(rho_hat-qnorm(0.975)*SE,rho_hat+qnorm(0.975)*SE); CI
## CI [1] 0.8107681 0.8492319

## Checking results:
## Pearson correlation and "normal" confidence intervals.
install.packages("confintr")
library(confintr)
ci_cor(X,Y)
#Sample estimate: 0.83 
#Confidence interval:
#    2.5%     97.5% 
#    0.8096678 0.8483423 

## Also bootstrap confidence intervals are supported and are the only option for rank correlations. 
# install.packages('boot')
library(boot)
ci_cor(X,Y, method = "pearson", type = "bootstrap", R = 1000, seed = 1)

# Sample estimate: 0.83 
# Confidence interval:
#       2.5%     97.5% 
#  0.8081769 0.8475616 
\end{lstlisting}

\subsection{Applications of the Delta-method in Regression Models}\label{rm}
To illustrate the use of the Delta and functional Delta-method based on the IF in multivariable regression settings we simulate data based on a cancer epidemiology example. Box 6 shows the data generating mechanism.\citep{ajph} Let's suppose that we want to emulate a clinical trial where we would like to estimate the effect of cancer treatment on the population average one-year risk of death standardized by age.  This hypothesis can be translated into an estimand -- specifically, the conditional RR, defined through a nonlinear combination of parameters, i.e., the ratio of two conditional probabilities. A regression estimator for the RR may use the predictions of a multivariate binomial regression model to compute the probabilities under different treatments. However, here we want to illustrate how to compute the SE for the RR. First, we fit the model with the binary indicator of one-year mortality as dependent variable and patient age and treatment status as independent variables. Then, from the fitted model and using the \textbf{predict} function we estimate the probability of death among cancer patients aged more than 65 years old had they been treated with mono-therapy versus the probability of death for the same patients had they been treated with dual therapy. Finally, we compute the conditional RR of death as the ratio between both probabilities using as reference those treated with dual therapy.
\\

\noindent \textbf{Box 6.} Data generation process: simulated example
\begin{lstlisting}
# Data generation
set.seed(1972)
N   <- 1000
# Age (1: > 65; 0: <= 65)
age <- rbinom(N,1,0.6)                                    
# Treatment (1: dual; 0=mono)
treat   <- rbinom(N,1,plogis(0.35 - 0.15*age)) 
# Counterfactual outcome under A=1 and A=0 respectively
death.1 <-  rbinom(N,1,plogis(2 - 1*1 + 0.65*age))
death.0 <-  rbinom(N,1,plogis(2 - 1*0 + 0.65*age))
# Observed outcome: mortality at 1 year after treatment initiation (1: death)
death   <- death.1*treat + death.0*(1 - treat)
# Mortality risk differences
mean(death.1-death.0)
## -0.13%
\end{lstlisting}

Our estimand is a nonlinear function of the regression coefficients from a logistic regression model of the one-year probability of death, given patients' age and treatment status. The estimand can be defined as follows:

\begin{equation*}
    \mathbb{P}(Y = 1 | X = x) = \dfrac{1}{1 + e^{-(\beta_0 + \sum_{i=1}^k \beta_i x_i)}}
\end{equation*}
where $Y$ refers to death within a year from study entry, and $X$ refers to the vector of covariates included in the model, i.e., age and cancer treatment. 

In the previous equation, the parameter vector $\beta$ has length $k + 1$ where its components are $\beta_{0}$,  the intercept term, and $\beta_{1}$ and $\beta_{2}$ representing the coefficients related to patients' age ($x_1$) and treatment status ($x_2$). We define  $x_1 = 0$ to represent age $< 65$ years and $x_1 = 1$ age $\geq 65$ years. Similarly, we have $x_2 = 0$ if patients are treated with monotherapy, while $x_2 = 1$ stands for dual therapy. Untreated individuals are not included in this example. For instance, the probability of one-year mortality among cancer patients aged more than 65 years old  ($x_1 = 1$) and treated with mono-therapy($x_2 = 0$) is computed as follows: 
\begin{equation*}
    \mathbb{P}\big(Y = 1 | x = (1,0)\big) = \dfrac{1}{1 + e^{-(\beta_0 + \beta_1 \cdot 1 + \beta_2 \cdot 0)}},
\end{equation*}
On the other hand, the probability of one-year mortality for those cancer patients aged more than 65 years old treated with dual therapy is given by the following expression:
\begin{equation*}
    \mathbb{P}\big(Y = 1 | x = (1,1)\big) = \dfrac{1}{1 + e^{-(\beta_0 + \beta_1 \cdot 1 + \beta_2 \cdot 1)}}.
\end{equation*}
Our estimand, the relative risk between those two groups is a function of the previous two probabilities and hence results in a function of the $beta$s: $\phi(\beta_0, \beta_1, \beta_2)$.
\begin{equation*}
\phi(\beta_0, \beta_1, \beta_2) = \dfrac{\mathbb{P}\big(Y = 1 | x = (1,1)\big)}{\mathbb{P}\big(Y = 1 | x = (1,0)\big)} = \dfrac{1 + e^{-(\beta_0 + \beta_1 + \beta_2)}}{1 + e^{-(\beta_0 + \beta_1)}}.
\end{equation*}

The influence function (i.e., steps(\ref{step1}, \ref{step2}) is computed as the product of the gradient of $\phi(\beta)$ with the direction vector $v:= \hat{\beta} - \beta$. So, using the principles of equation \eqref{taylorhadamard}, we take the partial derivatives to obtain the gradient: 

\begin{equation*}
    \nabla \phi (\beta_0, \beta_1, \beta_2) = \frac{- e^{- \beta_{2}}}{(e^{\beta_{0} + \beta_{1}} + 1)^2} \left[\begin{matrix}\left(1 - e^{\beta_{2}}\right) e^{\beta_{0} + \beta_{1}}, & \left(1 - e^{\beta_{2}}\right) e^{\beta_{0} + \beta_{1}}, &  e^{\beta_{0} + \beta_{1}} + 1 \end{matrix}\right]^T
\end{equation*}
where taking the direction vector $v = (\hat\beta_0 - \beta_0, \hat\beta_1 - \beta_1, \hat\beta_2 - \beta_2)^T$ we can compute the influence function (Hadamard derivative) as:
\begin{equation}
    IF(\beta) = \nabla   \phi (\beta_0, \beta_1, \beta_2)^T v = \frac{e^{- \beta_{2}}}{(e^{\beta_{0} + \beta_{1}} + 1)^2} \left(\left( \beta_{2} - \hat{\beta}_{2}\right) \left(e^{\beta_{0} + \beta_{1}} + 1\right) + \left(1 - e^{\beta_{2}}\right) e^{\beta_{0} + \beta_{1}} \left( \beta_{0} - \hat{\beta}_{0} + \beta_{1} - \hat{\beta}_{1} \right)\right) . 
\end{equation}

An expression for the variance is given by:
\begin{equation}
\begin{aligned}
    \text{Var}\big[IF(\beta)\big] & = \frac{e^{-2\beta_2}}{\left(e^{\beta_{0} + \beta_{1}} + 1\right)^{4}} \bigg[ \left(e^{\beta_{0} + \beta_{1}} + 1\right)^2 \cdot \text{Var}[\hat\beta_2] +  \left(1 - e^{\beta_{2}}\right)^2 e^{2(\beta_{0} + \beta_{1})} \cdot \left( \textrm{Var}\big[\hat\beta_0\big] + \textrm{Var}\big[\hat\beta_1\big] + 2\textrm{Cov}\big(\hat\beta_0,\hat\beta_1\big)\right)
    \\ & \qquad + 2\left(1 - e^{\beta_{2}}\right) e^{\beta_{0} + \beta_{1}} \left(e^{\beta_{0} + \beta_{1}} + 1\right)\Big( \textrm{Cov}\big(\hat\beta_0,\hat\beta_2\big)
    + \textrm{Cov}\big(\hat\beta_1,\hat\beta_2\big)\Big)
    \bigg]. 
\end{aligned}    
\end{equation}

Notice that the previous expression can be written (and is equivalent to) a matrix product. This can be seen by using the fact that maximum likelihood estimator $\hat{\beta}$ of $\beta$ is asymptotically normal. It has mean $\beta = (\beta_0, \beta_1, \beta_2)^T$ and covariance matrix (obtained from Fisher's information) given by\cite{izenman2008modern}:
$$
\Sigma_W =  \begin{pmatrix}
\text{Var}[\hat\beta_0] & \text{Cov}(\hat\beta_0, \hat\beta_1) & \text{Cov}(\hat\beta_0, \hat\beta_2) \\
\text{Cov}(\hat\beta_0, \hat\beta_1) & \text{Var}[\hat\beta_1] & \text{Cov}(\hat\beta_1, \hat\beta_2) \\
\text{Cov}(\hat\beta_0, \hat\beta_2) & \text{Cov}(\hat\beta_1, \hat\beta_2) & \text{Var}[\hat\beta_2]
\end{pmatrix}
$$
in conjunction with the fact that if $\hat{\beta} - \beta$ converges to a variable $Z$ with normal distribution then $\phi(\hat\beta) - \phi(\beta)$ converges to $\nabla \phi (\beta)\cdot  Z$ (see \eqref{approxTaylorv3}) to obtain an expression for the variance: 
\begin{equation}
\begin{aligned}
 \textrm{Var}\big[\phi(\hat\beta) - \phi(\beta)\big] & \approx \textrm{Var}\big[\nabla \phi (\beta)^T Z\big] = \nabla \phi (\beta) \textrm{Var}\big[Z \big]  \nabla \phi (\beta)^T =\nabla \phi (\beta) \Sigma_W  \nabla \phi (\beta)^T 
\end{aligned}
\end{equation}
where the last term of the equality is obtained from the fact that for a random $n$-dimensional vector $V$ and $a\in\mathbb{R}^n$:
$$
\textrm{Var}[ a V] = a \textrm{Var}[V] a^T.
$$

In Box 7 we show how to compute the SE using the Delta-method for the conditional RR derived from the marginal probabilities of the generalized linear model with the binary indicator of one-year mortality as dependent variable and patients age and treatment status as independent variables. Furthermore, we check the consistency of our results using the Delta-method implemented on the Multi-State Markov and Hidden Markov Models in Continuous Time R-package (MSM).\citep{Jackson2011Multi-StateR}

\[\]
\noindent \textbf{Box 7.} Delta-method to derive the SE for the conditional RR
\begin{lstlisting}
data  <- as.data.frame(cbind(death , treat , age))
m     <- glm(death ~ age + treat, family = binomial, data = data)
pMono <- predict(m, newdata = data.frame(age = 1, treat = 0), type = "response")
pDual <- predict(m, newdata = data.frame(age = 0, treat = 1), type = "response")
rr <- pMono / pDual
cat("Conditional risk ratio: ", rr)
# Conditional Risk Ratio:  1.330238

# The partial derivative are computed in R as follows:
x1 <- 1
x2 <- 0
x3 <- 0
x4 <- 1
b0 <- coef(m)[1]
b1 <- coef(m)[2]
b2 <- coef(m)[3]
e1 <- exp(- b0 - 1*b1 - 0*b2)
e2 <- exp(- b0 - 0*b1 - 1*b2)
p1 <- 1 / (1 + e1)
p2 <- 1 / (1 + e2)
# check rr 
p1/p2
# 1.330238
dfdb0 <- -e2*p1 + (1 + e2)*p1*(1 - p1)
dfdb1 <- -x2*e2*p1 + (1 + e2)*x1*p1*(1 - p1)
dfdb2 <- -x4*e2*p1 + (1 + e2)*x3*p1*(1 - p1)
grad <- c(dfdb0, dfdb1, dfdb2)
var <- t(grad) %*% vcov(m) %*% (grad)
se_rr <- c(sqrt(var))
se_rr
# 0.06057779
# Check with implemented delta-method in library msm 
library(msm)
se_rr_delta <- deltamethod( ~(1 + exp(-x1 -0*x2 -1*x3)) /
                                (1 + exp(-x1 -1*x2 -0*x3)), 
                            c(b0, b1, b2), 
                            vcov(m));se_rr_delta
## 0.06057779
# We obtain the same results for the SE of the RR computed before

# Finally, we compute the type Wald 95% CI
lb <- rr - qnorm(.975) * sqrt(vG)
ub <- rr + qnorm(.975) * sqrt(vG)
# Conditional Risk Ratio (95%CI)
c(lb, ub)
##  1.211508 1.448968
\end{lstlisting}

\section{Conclusion} 
During recent years new methods and estimators have been and continue to be developed to be applied in observational studies. For these new types of problems there is not an immediately available closed analytical form to derive the SE, complicating inference. Yet, the distribution of the statistic can be approximated based on asymptotic theory to directly estimate its variance and hence the SE. In this tutorial, we provided an introduction to the classic and functional Delta-method and linked it to the IF. We facilitated the approach by proposing a three-step procedure to estimate the SE of functionals using the first order Taylor expansion and the Hadamard derivative. These steps are needed to analytically derive the IF and compute its variance. Furthermore, we provided  R code to apply this approach for statistical examples and one in cancer epidemiology research.

The Delta-method is widely used  for statistical inference with classical epidemiological methods to estimate e.g. associational measures. Furthermore, it is used to derive the SE of an asymptotically linear estimator or functions that can be approximated by averages, such as the conditional estimations of the parameters fitted in regression models. It is closely related to M-estimation; see, for example, the Huber Sandwich estimator of the variance.\citep{deMenezes2021AAnalysis,Freedman2006,Efron1993} Therefore, the application of the Delta-method relies on asymptotic normality which in some cases can be a strong assumption. For example, when functions are not analytical (\textit{i.e.} do not converge to their Taylor series), such as those described in section \ref{counterexample}, or are sparse enough relative to the dimension, and in the case of non-differentiability, the Delta-method cannot be used to derive the SE of the statistic.\citep{Kennedy2022SemiparametricReview} 

There are more conservative approaches available for statistical inference based on re-sampling such as the bootstrap.\citep{Efron1993} Using the Bootstrap, the original sample approximates the population from which it was drawn. The bootstrap distribution of a statistic, based on many re-samples, approximates the sampling distribution of the statistic. Thus, boostrapping can only be applied under certain smoothness conditions,\citep{efron1982, efron1983} and it cannot be applied when using data-adaptive estimation procedures. However, as shown in this article, the functional Delta-method based on the IF can be used instead to derive the variance of the statistic via the functional Delta-method.\\

Currently, there is an active research community developing methods and tools for high dimensional data analysis such as machine learning and causal inference. For example, within the causal inference field, the functional Delta-method and the IF are used to develop plug-in variance estimators for data-adaptive doubly robust causal inference methods.\citep{vanderLaan2021HigherEstimation, Kennedy2022SemiparametricReview}. Due to the computational complexity of some machine learning methods, for many new techniques of causal inference using data-adaptive procedures, the functional Delta-method based on the IF will be preferred over the Bootstrap for statistical inference. The increasing  interest in these methods justifies the need for applied tutorials to further disseminate their use and application.\citep{Smith} Furthermore, the derivation of the IF for these new methods needs to be introduced and taught among applied statisticians. New methodological focuses and emerging methods requires that applied statisticians and methods-focused epidemiologists extend their competencies in  statistical inference. We thus provided an overview of the Delta-method, the functional Delta-method, and the IF that may help to fill this gap.

\section*{Funding}
This work was supported by the Medical Research Council [grant number MR/W021021/1]. Miguel Angel Luque-Fernandez is supported by the Spanish National Institute of Health, Carlos III Miguel Servet I Investigator Award (CP17/00206). Mireille E. Schnitzer holds a Canada Research Chair from the Canadian Institutes of Health Research and a Discovery Grant from the Natural Sciences and Engineering Research Council of Canada. Aurelien Belot, Camille Maringe, and Bernard Rachet are supported by a Cancer Research UK Population Research Committee Programme Award (C7923/A29018). Michael Schomaker is supported by the German Research Foundations (DFG) Heisenberg Programm (grants 465412241 and 465412441).

\section*{Authors contributions}
The article arose from the motivation to disseminate the principles of modern epidemiology among clinicians and applied researchers. MALF developed the concept, and wrote the first drat of the article. RZT and MALF wrote further versions of the article. All authors interpreted and reviewed the code and the data, drafted and revised the manuscript. All authors read and approved the final version of the manuscript. MALF is the guarantor of the article.

\section*{Acknowledgments}
The motivation and some parts of the tutorial come from MALF's work in a visiting academic position in the Division of Biostatistics at the Berkeley School of Public Health in 2019. We would like to acknowledge Professor Mark van der Laan's effort in teaching and disseminating modern statistical methods applied to causal inference. We thank professors Ashley Naimi, Sho Kumakai, and Antoine Chambaz for their comments on  earlier versions of the manuscript. 

This research was funded, in part, by the Medical Research Council [grant number MR/W021021/1]. A CC BY or equivalent licence is applied to the Author Accepted Manuscript (AAM) arising from this submission, in accordance with the grant’s open access conditions.

\bibliography{bibfile.bib, extrabibliography.bib}

\end{document}